\documentclass[12pt]{article}
\usepackage{a4wide}
\usepackage{latexsym}
\usepackage{amsmath}
\usepackage{amsfonts}
\usepackage{amscd}
\usepackage{amsthm}
\usepackage{amssymb}
\usepackage{cite}
\usepackage{shuffle}
\usepackage{hyperref}
\usepackage{pslatex}
\usepackage{graphicx}
\usepackage[latin1,utf8]{inputenc}
\usepackage[T1]{fontenc}
\usepackage{euscript}
\usepackage{subfigure}
\usepackage{bm}

\usepackage{mathtools}
\usepackage{arydshln}

\usepackage{color}

\newcommand{\bq}{\begin{eqnarray}}
\newcommand{\eq}{\end{eqnarray}}

\theoremstyle{plain}

\usepackage{color}

\allowdisplaybreaks

\begin{document}

\thispagestyle{empty}

\begin{flushright}
TUM-HEP-1456/23
\end{flushright}

\vspace{1.5cm}

\begin{center}
  {\Large\bf $\varepsilon$-factorized differential equations for two-loop non-planar triangle Feynman integrals with elliptic curves\\
  }
  \vspace{1cm}
  {\large Xuhang  Jiang${}^{a}$, Xing Wang${}^{b}$, Li Lin Yang${}^{c}$ and Jingbang Zhao${}^{a}$\\
  \vspace{1cm}
      {\small \em ${}^{a}$ School of Physics and State Key Laboratory of Nuclear Physics and Technology, } \\
      {\small \em Peking University, Beijing 100871, China}\\
  \vspace{2mm}
      {\small \em ${}^{b}$  Technische Universit\"at M\"unchen, TUM School of Natural Sciences, } \\
      {\small \em Physik Department, 85748 Garching, Germany}\\
  \vspace{2mm}
       {\small \em ${}^{c}$ Zhejiang Institute of Modern Physics, School of Physics,} \\
      {\small \em Zhejiang University, Hangzhou 310027, China}

  }
\end{center}

\vspace{2cm}

\begin{abstract}
This paper investigates two-loop non-planar triangle Feynman integrals involving elliptic curves. In contrast to the Sunrise and Banana integral families, the triangle families involve non-trivial sub-sectors. We show that the methodology developed in the context of Banana integrals can also be extended to these cases and obtain $\varepsilon$-factorized differential equations for all sectors. The letters are combinations of modular forms on the corresponding elliptic curves and algebraic functions arising from the sub-sectors. With uniform transcendental boundary conditions, we express our results in terms of iterated integrals order-by-order in the dimensional regulator, which can be evaluated efficiently. Our method can be straightforwardly generalized to other elliptic integral families and has important applications to precision physics at current and future high-energy colliders.
\end{abstract}

\vspace*{\fill}

\newpage

\section{Introduction}
\label{sect:intro}

Precision predictions for observables in quantum field theories are crucial for extracting information from experimental data at facilities such as the CERN Large Hadron Collider (LHC). To achieve this, one needs to calculate scattering amplitudes at the multi-loop level, which boils down to calculating corresponding Feynman integrals. The first step in this process is to reduce all relevant Feynman integrals to linear combinations of a smaller set of master integrals (MIs). There are several ways to perform this reduction, with the most commonly used being the integration-by-parts (IBP) technique with the Laporta algorithm~\cite{Laporta:2000dsw}. Several public packages implementing this algorithm are available, e.g., \texttt{Litered}~\cite{Lee:2012cn, Lee:2013mka}, \texttt{FIRE}~\cite{Smirnov:2019qkx}, \texttt{Reduze}~\cite{vonManteuffel:2012np} and \texttt{Kira}~\cite{Klappert:2020nbg}. However, the IBP systems can become prohibitively expensive to solve for cutting-edge applications. In recent years, novel approaches have emerged to bypass these cumbersome linear systems, e.g., the GKZ system~\cite{nasrollahpoursamami2016periods, Vanhove:2018mto, delaCruz:2019skx} and the intersection theory~\cite{Mastrolia:2018uzb, Frellesvig:2019uqt}, just to name a few.

After reduction, a system of differential equations can be written down for the master integrals with respect to kinematic variables or some auxiliary variables~\cite{Kotikov:1990kg, Kotikov:1991pm, Remiddi:1997ny, Gehrmann:1999as, Liu:2017jxz}. With suitable boundary conditions, this system can always be solved numerically by direct integration or by series expansions \cite{Hidding:2020ytt, Liu:2020kpc, Liu:2022chg}. It is conceptually more interesting to solve the differential equations analytically, which is, however not an easy task in general.
In dimensional regularization, with the spacetime dimension $D=4-2\varepsilon$, the differential equations get simpler if the dependence on $\varepsilon$ is entirely factorized out~\cite{Henn:2013pwa}. In this case, the solutions can be obtained order-by-order in $\varepsilon$ as iterated integrals, which often lead to analytic expressions in terms of multiple polylogarithms (MPLs)~\cite{Goncharov:1998kja, Goncharov:2001iea}. Hence, the problem of analytically calculating Feynman integrals often reduces to finding appropriate transformations to turn the differential equations into this kind of $\varepsilon$-form.

For an integral family involving only logarithmic singularities, it is generally believed that a system of differential equations in the $\varepsilon$-form always exists. Various methods to find it have been proposed in the literature \cite{Argeri:2014qva, Gehrmann:2014bfa, Lee:2014ioa, Adams:2017tga, Lee:2017oca, Dlapa:2020cwj, Gituliar:2017vzm, Prausa:2017ltv, Meyer:2017joq, Lee:2020zfb, Chen:2020uyk, Chen:2022lzr}. For more generic integrals, it is unclear whether the factorization of $\varepsilon$-dependence can always be achieved. For integrals involving elliptic curves or even beyond, there is growing evidence that this statement may be true. In particular, this has been shown for equal-mass Banana integrals at arbitrary loop order~\cite{Pogel:2022yat, Pogel:2022ken, Pogel:2022vat}. In this paper, we provide more examples to support this statement using the geometry-inspired method for Banana integrals~\cite{Adams:2015ydq, Pogel:2022yat, Pogel:2022ken, Pogel:2022vat}.

Our understanding of Feynman integrals and scattering amplitudes benefits greatly from the geometric point of view. The simplest class of functions, the MPLs, are related to Abelian differentials of the first kind on a Riemann sphere (a genus-0 object). Their integrands are rational functions, and can usually be cast into the $d\log$-form \cite{Arkani-Hamed:2012zlh,Bern:2014kca,Herrmann:2019upk,Henn:2020lye,Chen:2020uyk}.  The next-to-simplest class of functions are related to elliptic curves, which have received much interest in recent decades~\cite{Adams:2017ejb,Broedel:2017kkb,Adams:2018bsn,Bogner:2019lfa,Giroux:2022wav,Sogaard:2014jla,Bonciani:2016qxi,Ablinger:2017bjx,Broedel:2019hyg,Bourjaily:2017bsb,Kristensson:2021ani,Abreu:2019fgk,Kniehl:2019vwr}. These elliptic-curve related functions appear in many cutting-edge calculations and are of crucial importance for precision predictions at future colliders.

An elliptic Feynman integral family may be associated with one or more elliptic curves (an example with two curves contributes to top quark pair production at the LHC and was studied in \cite{Muller:2022gec}). In this paper, we focus on integral families involving only one elliptic curve. Given an integral family, the associated elliptic curve can be obtained from the maximal cut in the Baikov representation \cite{Baikov:1996iu, Lee:2010wea} or from the variety of the second graph polynomial (see \cite{Weinzierl:2022eaz} for a comprehensive review). The curve is generically parameterized by external variables such as momentum scalar products and masses. In this paper, we consider elliptic curves depending on only one dimensionless parameter. This can happen if the integrals involve only two physical scales or if one constrains to a two-dimensional surface in a higher-dimensional parameter space. In these single-parameter cases, evidence shows that the best variable to use is the modular variable $\tau$ defined by the ratio of the two periods of the underlying elliptic curve, which has been employed to cast the differential equations into the $\varepsilon$-form for Sunrise and Banana families~\cite{Adams:2015ydq, Pogel:2022yat, Pogel:2022ken, Pogel:2022vat}.

\begin{figure}[t!]
  \centering
  \subfigure[]
     {
      \label{subfig:diagrama}
      \includegraphics[width=7cm]{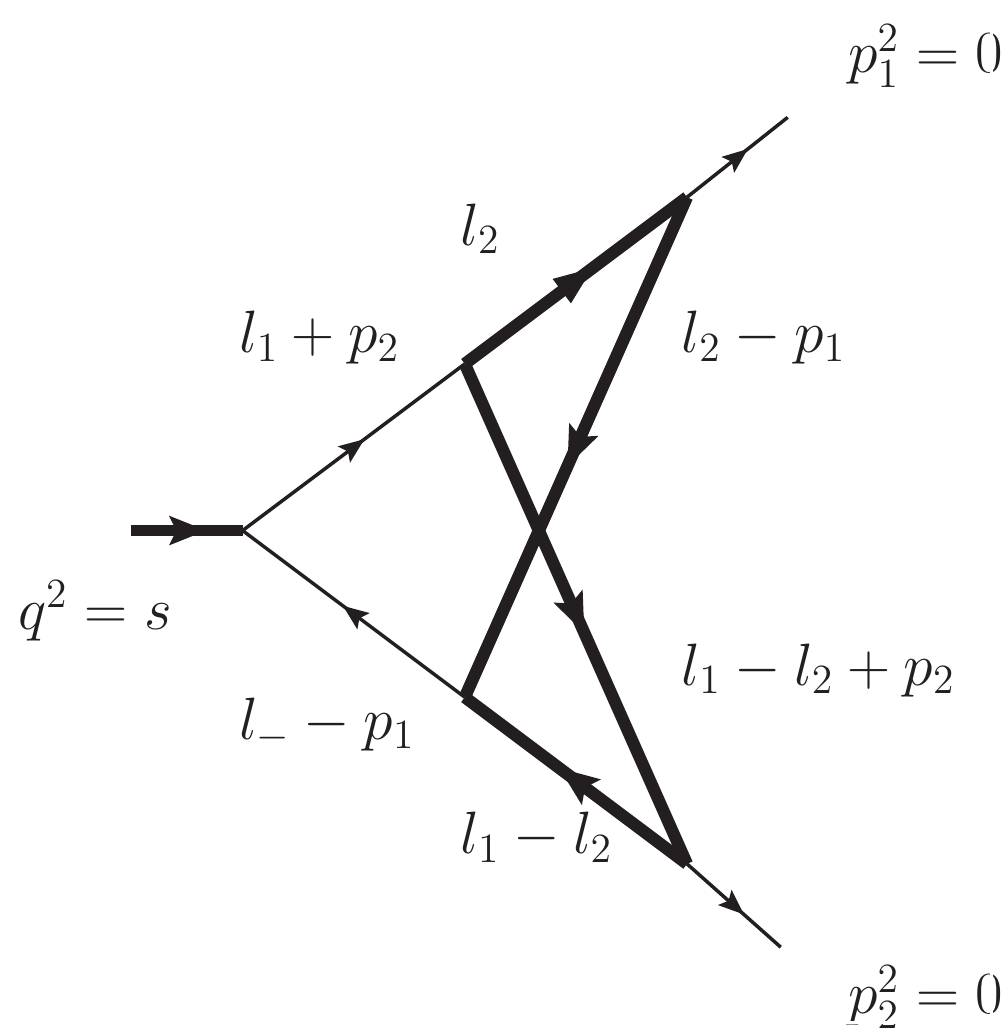}
     }
   \hspace{2em}
   \subfigure[]
     {
      \label{subfig:diagramb}
      \includegraphics[width=7cm]{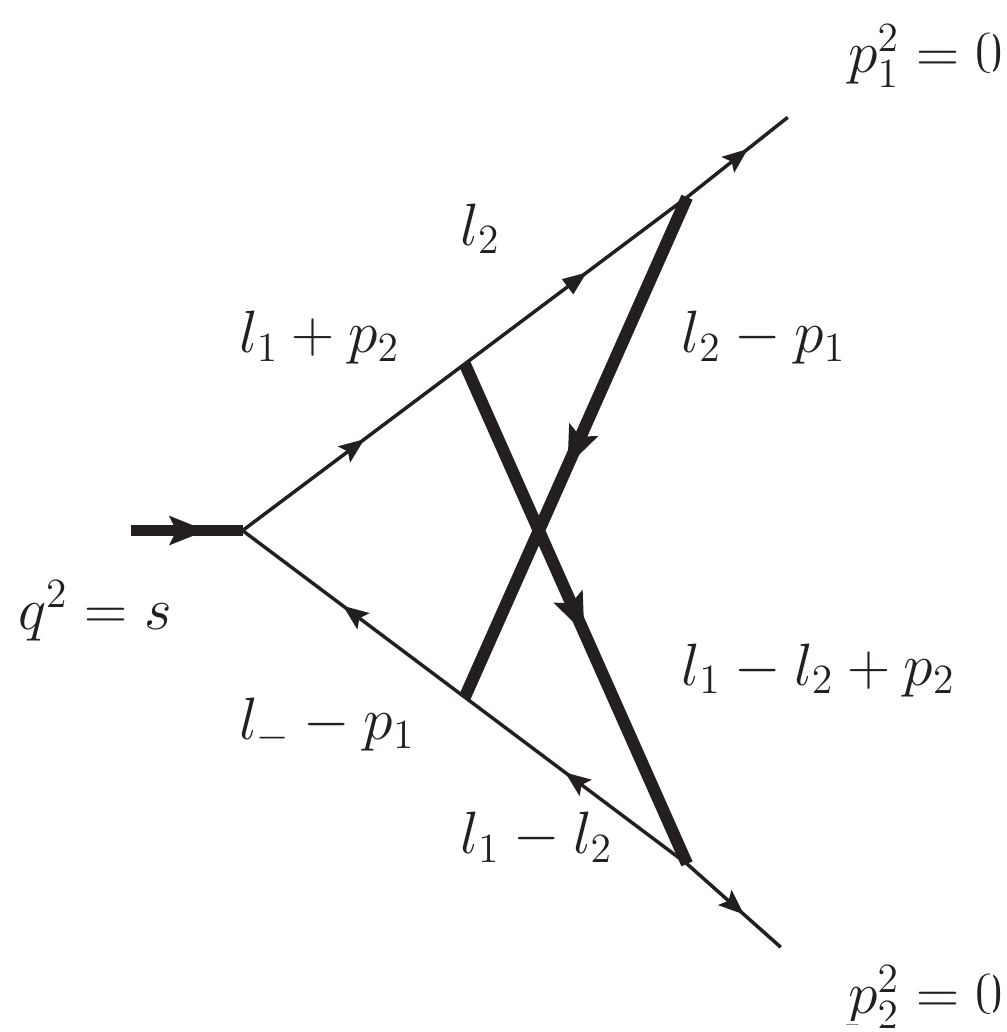}
      }
  \caption{\label{fig:NPTri}Two non-planar triangle integral families at two loops. Family (a) (the left diagram) involves a closed loop with the same mass and was studied in \cite{vonManteuffel:2017hms}. Family (b) (the right diagram) involves two massive propagators.}
\end{figure}

In this paper, we revisit two non-planar triangle integral families from the above perspective. The diagrams for the two families are shown in Fig.~\ref{fig:NPTri}. Compared to the Sunrise and Banana families, these triangle families involve non-trivial sub-sectors. It is hence interesting to investigate whether the method developed in \cite{Adams:2015ydq, Pogel:2022yat, Pogel:2022ken, Pogel:2022vat} can be extended to find $\varepsilon$-factorized differential equations in these cases, and to express the solutions as well-behaved iterated integrals with nice analytic and algebraic properties, that can be easily evaluated numerically. Besides being theoretically interesting, these integrals are also of phenomenological importance. The family (a) appears, e.g., in bottom-quark loop contributions to the $H \to gg$ decay, in top-quark loop contributions to the di-photon and di-jet productions at the LHC, as well as in the pair production and associated production of the Higgs boson under suitable limits. The family (b) appears, e.g., in the next-to-next-to-leading order (NNLO) electroweak (EW) corrections to the $H \to b\bar{b}$ decay and to the di-jet production. The investigation in this paper, therefore, serves as a step towards a fully analytic understanding of these important scattering amplitudes.

The paper is organized as follows. In Section~\ref{sect:setup} we set up our notation and present the results for the sub-sector integrals in both family (a) and (b). For the elliptic top-sector integrals, we first use the simpler case of family (a) in Section~\ref{sect:familya} to introduce our method and then apply it to the more complicated case of family (b) in Section~\ref{sect:familyb}. We conclude in Section~\ref{sect:conclusions}. All the results are provided in auxiliary files with the preprint.

\section{Setup and canonical sub-sector integrals}
\label{sect:setup}
We consider the integral families shown in Fig.~\ref{fig:NPTri}, where thick internal lines have the same mass $m$, and all the other internal lines are massless. The thick external leg has the offshellness $s = (p_1+p_2)^2$, and the other two external legs are massless $p_1^2=p_2^2=0$. In momentum space, the two integral families involve 6 propagators and an irreducible scalar product (ISP) $D_7 = l_1^2$. The integrals in the two families can be written as
\begin{equation}
	\label{eq:NPTriMomdef}
	\begin{aligned}
		I_{\nu_1\nu_2\cdots \nu_7} = e^{2\varepsilon\gamma_E}(m^2)^{\nu-D}\int\frac{d^D l_1}{i\pi^{D/2}}\int\frac{d^D l_2}{i\pi^{D/2}}\frac{D_7^{-\nu_7}}{D_1^{\nu_1} D_2^{\nu_2} D_3^{\nu_3} D_4^{\nu_4} D_5^{\nu_5} D_6^{\nu_6}} \,,
	\end{aligned}
\end{equation}
with $D=4-2\varepsilon$, $\nu=\nu_1+\cdots+\nu_6+\nu_7$. The propagator denominators are
\begin{equation}\label{eq:nptpd}
	\begin{aligned}
		&D_1 = (l_1-p_1)^2,\,\,D_2 = (l_2-p_1)^2-m^2,\,\,D_3=(l_1+p_2)^2,\,\,\\
		&D_4 = (l_1-l_2+p_2)^2-m^2,\,\,D_5 = (l_1-l_2)^2-\kappa\, m^2,\,\,D_6=l_2^2-\kappa\,m^2,
	\end{aligned}
\end{equation}
where $\kappa=1$ for family (a) and $\kappa=0$ for family (b). We have made the Feynman $i0$-prescription in the propagators implicit. With the pre-factor $(m^2)^{v-D}$, the integrals depend on only one dimensionless variable that we take as
\begin{equation}
	\label{eq:kinevar}
	y = -\frac{m^2}{s} \in \mathbb{R} + i 0 \, ,
\end{equation}
where the infinitesimal imaginary part is determined by the Feynman $i0$ prescription.

We use \texttt{Litered}~\cite{Lee:2013mka} and \texttt{Kira}~\cite{Klappert:2020nbg} to perform the IBP reductions. For family (a), there are 11 master integrals in total, 2 of which are in the top sector, while the others are in sub-sectors. For family (b), there are 18 master integrals in total, 3 of which are in the top sector, while the others are in sub-sectors. The pre-canonical master integrals for family (a) are given as
\begin{equation}
	\label{eq:prebasisa}
	\begin{aligned}
		\text{top}:\quad I^{(a)}_{1} &= I^{(a)}_{1111110}\,, \; I^{(a)}_{2} = I^{(a)}_{1111120}\,,
		\\
		\text{sub}:\quad I^{(a)}_{3} &= I^{(a)}_{1101110}\,, \; I^{(a)}_{4} = I^{(a)}_{1111100}\,, \; I^{(a)}_{5} = I^{(a)}_{0210110}\,, \; I^{(a)}_{6} = I^{(a)}_{1122000}\,, \; I^{(a)}_{7} = I^{(a)}_{1112000}\,,
		\\
		I^{(a)}_{8} &= I^{(a)}_{0220100}\,, \; I^{(a)}_{9} = I^{(a)}_{0210200} \,, \; I^{(a)}_{10} = I^{(a)}_{1220000} \,, \; I^{(a)}_{11} = I^{(a)}_{0202000} \, ,
	\end{aligned}
\end{equation}
and for family (b) as
\begin{equation}
	\label{eq:prebasisb}
	\begin{aligned}
		\text{top}:\quad I^{(b)}_{1} &= I^{(b)}_{1111110}\,, \; I^{(b)}_{2} = I^{(b)}_{1112110}\,, \; I^{(b)}_{3} = I^{(b)}_{1111120}\,,
		\\
		\text{sub}:\quad I^{(b)}_4 &= I^{(b)}_{1111100}\,, \; I^{(b)}_{5} = I^{(b)}_{1120110}\,, \; I^{(b)}_{6} = I^{(b)}_{1110110}\,, \; I^{(b)}_{7} = I^{(b)}_{1102110}\,, \; I^{(b)}_{8} = I^{(b)}_{1101110} \,,
		\\
		I^{(b)}_9 &= I^{(b)}_{1122000}\,, \; I^{(b)}_{10} = I^{(b)}_{1112000}\,, \; I^{(b)}_{11} = I^{(b)}_{0121200}\,, \; I^{(b)}_{12} = I^{(b)}_{0211200}\,, \; I^{(b)}_{13} = I^{(b)}_{1010120}\,,
		\\
		I^{(b)}_{14} &= I^{(b)}_{1220000}\,, \; I^{(b)}_{15} = I^{(b)}_{1200200}\,, \; I^{(b)}_{16} = I^{(b)}_{0220100}\,, \; I^{(b)}_{17} = I^{(b)}_{0120200}\,, \; I^{(b)}_{18} = I^{(b)}_{0202000} \, .
	\end{aligned}
\end{equation}
In this section, we first construct $\varepsilon$-factorized differential equations for the sub-sectors. Those for the top sectors will be discussed in the latter sections.

\subsection{Canonical basis for sub-sectors}
Neither of the sub-sectors in the two families involves elliptic curves. We use the method of \cite{Chen:2020uyk, Chen:2022lzr} to construct integrals with uniform transcendentality (UT) in the Baikov representations. Briefly speaking, we construct integrands in the generalized $\mathrm{d}\log$ forms:
\begin{equation}
	\left[P(\bm{z})\right]^{\epsilon}\mathrm{d} \log\alpha_{1}(\bm{z})\wedge \mathrm{d}\log\alpha_{2}(\bm{z})\wedge \cdots \wedge \mathrm{d}\log\alpha_{n}(\bm{z})\,,
\end{equation}
where $\bm{z}$ denotes the set of Baikov variables, which is simply a subset of the propagator denominators, i.e., $z_i \in \{D_i\}$. $P(\bm{z})$ is a rational function that defines the representation for the particular sector, and $\alpha_i(\bm{z})$ are algebraic functions of $\bm{z}$. As an example, in the sub-sector $\{1,1,1,0,1,1,0\}$ of family $(b)$, we can have the following construction
\begin{equation}
	\begin{aligned}
		\left[P(\bm{z})\right]^{\epsilon} \mathrm{d}\log z_{1}\wedge\mathrm{d}\log z_{2} \wedge\mathrm{d}\log z_{3}\wedge\mathrm{d}\log z_{5}\wedge\mathrm{d}\log z_{6}\wedge\mathrm{d}\log (z_{7}-z_{1}) \,.
	\end{aligned}
\end{equation}
The above integrand can be converted to a linear combination of Feynman integrals following the method presented in \cite{Chen:2020uyk, Chen:2022lzr}. With suitable normalization factors, it corresponds to the UT integral $\epsilon^4 \, x \, I^{(b)}_{1110110}$. Proceeding with each sub-sector, we arrive at the following 15 UT integrals in the sub-sectors of family (b):
\begin{equation}
    \label{eq:subUTb}
	\begin{aligned}
		M^{(b)}_{4} &= \varepsilon^4 x\,I^{(b)}_{1111100} \,,\quad M^{(b)}_{5} = \varepsilon^3 x^2\, I^{(b)}_{1120110} \,,\quad M^{(b)}_{6} = \varepsilon^4 x\, I^{(b)}_{1110110} \,,\quad M^{(b)}_{7} = \varepsilon^3 x\, I^{(b)}_{1102110}  \,, \\
		M^{(b)}_{8}&=\varepsilon^4 x\, I^{(b)}_{1101110} \,,\quad M^{(b)}_{9} = \varepsilon^2\sqrt{x(4+x)}\left(-x\,I^{(b)}_{1122000}+\frac{\varepsilon}{2(1+2\varepsilon)}I^{(b)}_{0202000}\right) , \\
		M^{(b)}_{10}&=\varepsilon^3 x\,I^{(b)}_{1112000} \,, \\
		M^{(b)}_{11}&=\varepsilon^2\frac{x}{2}\left((1+x)I^{(b)}_{0211200}-(1+x)I^{(b)}_{0121200}-2I^{(b)}_{0120200}+\frac{2-x}{x}I^{(b)}_{0202000}\right) ,\\
		M^{(b)}_{12}&=\varepsilon^2\frac{x}{4}\left((3+x)I^{(b)}_{0121200}+(1-x)I^{(b)}_{0211200}+2I^{(b)}_{0220100}+4I^{(b)}_{0120200}+I^{(b)}_{0202000}\right) ,\\
		M^{(b)}_{13}&=\varepsilon^3 x\,I^{(b)}_{1010120} \,,\quad M^{(b)}_{14} = \varepsilon^2 x\,I^{(b)}_{1220000} \,,\quad M^{(b)}_{15} = \frac{\varepsilon(\varepsilon-1)}{2}I^{(b)}_{1200200} \,,  \\
		M^{(b)}_{16}&=-\varepsilon^2\frac{x}{2} I^{(b)}_{0220100} \,,\quad M^{(b)}_{17} = -\varepsilon^2\frac{1-x}{4}\left(I^{(b)}_{0120200}+2I^{(b)}_{0220100}\right) ,\quad M^{(b)}_{18} = \varepsilon^2 I^{(b)}_{0202000} \,,
	\end{aligned}
\end{equation}
where $x \equiv -1/y = s/m^2$. The differential equations for this UT basis take the $\varepsilon$-form (hereafter, we will suppress the superscript (a) or (b) when it is clear from the context):
\begin{align}\label{eq:subepsb}
		\frac{\mathrm{d}}{\mathrm{d}y} M_{4}&= \frac{\varepsilon}{y} \left(-2M_{10}+\frac{3}{2}M_{11}+M_{12}+M_{16}-\frac{3}{2}M_{18}\right) , \nonumber \\
		\frac{\mathrm{d}}{\mathrm{d}y} M_{5}&= \frac{\varepsilon}{y(1+y)}\left(2M_{5}-2M_{6}-2M_{13}+M_{14}-M_{15}-2M_{16}-2M_{17}\right) , \nonumber \\
		\frac{\mathrm{d}}{\mathrm{d}y} M_{6}&= \frac{\varepsilon}{y}\left(M_{5}-M_{6}-\frac{1}{2}M_{15}-3M_{16}-M_{17}\right) , \nonumber \\
		\frac{\mathrm{d}}{\mathrm{d}y} M_{7}&=\varepsilon \left(-\frac{1+3y}{y(1+y)}M_{7}-\frac{2}{1+y}M_{8}+\frac{2+3y}{y(1+y)}M_{11}+\frac{2}{y}M_{12}-\frac{1}{y(1+y)}M_{15}\right. \nonumber \\
		&\hspace{4em} \left.+\frac{2}{y}M_{16}-\frac{5+6y}{2y(1+y)}M_{18}\right) , \nonumber \\
		\frac{\mathrm{d}}{\mathrm{d}y} M_{8}&= \frac{\varepsilon}{y}\left(M_{7}-\frac{3}{2}M_{11}-M_{12}-M_{16}+\frac{3}{2}M_{18}\right) , \nonumber \\
		\frac{\mathrm{d}}{\mathrm{d}y} M_{9}&=\varepsilon \left(\frac{3-4y}{y(1-4y)}M_{9}-\frac{3}{y\sqrt{1-4y}}M_{10}+\frac{1}{y\sqrt{1-4y}}M_{14}+\frac{1}{y\sqrt{1-4y}}M_{18}\right) , \nonumber \\
		\frac{\mathrm{d}}{\mathrm{d}y} M_{10}&=\varepsilon \left(\frac{1}{y\sqrt{1-4y}}M_{9}-\frac{1}{y}M_{10}\right) , \nonumber \\
		\frac{\mathrm{d}}{\mathrm{d}y} M_{11}&= \frac{\varepsilon}{y}\left(-M_{11}-2M_{12}+M_{18}\right) , \nonumber \\
		\frac{\mathrm{d}}{\mathrm{d}y} M_{12}&=\varepsilon \left(\frac{1}{y(1-y)}M_{11}+\frac{2}{y}M_{12}+\frac{4}{y(1-y)}M_{16}+\frac{4}{y(1-y)(1+y)}M_{17}\right. \nonumber \\
		&\hspace{4em} \left. -\frac{3}{2y(1-y)}M_{18}\right) , \nonumber \\
		\frac{\mathrm{d}}{\mathrm{d}y} M_{13}&= \frac{2\varepsilon}{y}M_{13} \, , \quad \frac{\mathrm{d}}{\mathrm{d}y} M_{14} = \frac{\varepsilon}{y} \, M_{14} \, , \quad \frac{\mathrm{d}}{\mathrm{d}y} M_{15}=0 \, , \nonumber \\
		\frac{\mathrm{d}}{\mathrm{d}y} M_{16}&=\varepsilon \left(-\frac{1}{y}M_{16}-\frac{2}{y(1+y)}M_{17}\right) , \nonumber \\
		\frac{\mathrm{d}}{\mathrm{d}y} M_{17}&=\varepsilon \left(\frac{3}{y}M_{16}+\frac{4}{y(1+y)}M_{17}\right) , \quad  \frac{\mathrm{d}}{\mathrm{d}y} M_{18}=0 \, .
\end{align}

The construction of a UT basis for the family (a) can be similarly done. For ease of comparison with the literature, we adopt the existing basis of \cite{vonManteuffel:2017hms}, see also\footnote{We thank Christoph Nega and Lorenzo Tancredi for sharing preliminary results for related topics.} \cite{TUM:2023eps}:
\begin{equation}
    \label{eq:subUTa}
	\begin{aligned}
		M^{(a)}_{3} &= \varepsilon^4 x\,I^{(a)}_{0111110} \,, \quad M^{(a)}_{4} = \varepsilon^4 x\,I^{(a)}_{1011110} \,, \\
		M^{(a)}_{5} &= \varepsilon^2\sqrt{x(4+x)}\left(x\,I^{(a)}_{1122000}-\frac{\varepsilon}{2(1+2\varepsilon)}I^{(a)}_{0202000}\right) , \\		
		M^{(a)}_{6} &= \varepsilon^3 x\,I^{(a)}_{1010210}, \quad M^{(a)}_{7} = \varepsilon^3 x\,I^{(a)}_{0211100},\quad M^{(a)}_{8} = \varepsilon^2 x\,I^{(a)}_{0210200}, \\
		M^{(a)}_{9} &= \varepsilon^2\sqrt{x(x-4)}\left( \frac{1}{2}I^{(a)}_{0210200} + I^{(a)}_{0220100}\right) \,, \\
		M^{(a)}_{10} &= \varepsilon^2 x\,I^{(a)}_{1020020} \,, \quad M^{(a)}_{11} = \varepsilon^2 I^{(a)}_{0000220} \,.
	\end{aligned}
\end{equation}
The $\varepsilon$-form differential equations for the sub-sectors are given by:
\begin{equation}
	\label{eq:epssuba}
		 \frac{d}{d y}\begin{pmatrix}
			M_3\\
			M_4\\
			M_5\\
			M_6\\
			M_7\\
			M_8\\
			M_9\\
			M_{10}\\
			M_{11}
		\end{pmatrix} = \frac{\varepsilon}{y}
		\begin{pmatrix}
		\quad 0\quad & \quad 0\quad & \quad 0\quad & \quad 0\quad & \quad -1\quad & \quad 0\quad & \quad 0\quad & \quad 0\quad & \quad 0\quad \\
		0 & 1 & 0 & -1 & \frac{1}{2} & 0 & 0 & 0 & 0 \\
		0 & 0 & \frac{4 y-3}{4 y-1} & \frac{3}{\sqrt{1-4 y}} & 0 & 0 & 0 & \frac{-1}{\sqrt{1-4 y}} & \frac{-1}{\sqrt{1-4 y}} \\
		0 & 0 & \frac{-1}{\sqrt{1-4 y}} & -1 & 0 & 0 & 0 & 0 & 0 \\
		0 & 0 & 0 & 0 & 0 & -1 & 0 & 0 & 0 \\
		0 & 0 & 0 & 0 & 0 & -1 & \frac{-2}{\sqrt{4 y+1}} & 0 & 0 \\
		0 & 0 & 0 & 0 & 0 & \frac{3}{\sqrt{4 y+1}} & \frac{4 (y+1)}{4 y+1} & 0 & \frac{-1}{\sqrt{4 y+1}} \\
		0 & 0 & 0 & 0 & 0 & 0 & 0 & 1 & 0 \\
		0 & 0 & 0 & 0 & 0 & 0 & 0 & 0 & 0
		\end{pmatrix}
		\begin{pmatrix}
			M_3\\
			M_4\\
			M_5\\
			M_6\\
			M_7\\
			M_8\\
			M_9\\
			M_{10}\\
			M_{11}
		\end{pmatrix}.
\end{equation}

\subsection{Iterated integrals}
\label{sect:iter}

The solutions to the $\varepsilon$-factorized differential equations such as Eqs.~\eqref{eq:subepsb} and \eqref{eq:epssuba} can be naturally represented by iterated integrals~\cite{Chen:1977oja} combined with suitable boundary conditions. Here we summarize the definitions that we will need in our calculations and refer the readers to \cite{Weinzierl:2022eaz} for a systematic review. Let $f_i(z)$'s be functions of $z$, called letters. We define the $n$-fold iterated integral of these letters by:
\begin{equation}
	\label{eq:itedef}
	I(f_{1}, f_{2}, \ldots, f_{n} ; \, z_{0}, \, z) = \int_{z_{0}}^{z} d z_{1}\,f_{1}(z_{1}) \int_{z_{0}}^{z_{1}} d z_{2}\,f_{2}(z_{2}) \ldots \int_{z_{0}}^{z_{n-1}} d z_{n}\, f_{n}(z_{n}) \, ,
\end{equation}
where $z_0$ is the base point. By convention, one sets $I(;\, z_0, z) = 1$. When the integration develops logarithmic singularity for $z_i\to z_0$, the standard ``trailing zero'' or ``tangential base point'' regularization \cite{brown2017multiple, Adams:2017ejb} is adopted. Acting on an iterated integral, the symbol map \cite{Goncharov:2010jf} simply picks up the sequence of its letters:
\begin{equation}
\mathcal{S}[I(f_{1}, f_{2}, \ldots, f_{n} ; \, z_{0}, \, z)] = f_n(z) \otimes \cdots \otimes f_2(z) \otimes f_1(z) \, .
\end{equation}
For the sub-sector integrals of family (a) and (b), the asymptotic boundary conditions at $y=0$ can be easily obtained and are collected in the Appendix. Combining the boundary conditions and the iterated integrals from the differential equations, we obtain the solutions for Eqs.~\eqref{eq:subepsb} and \eqref{eq:epssuba} and give them in the electronic files attached to this paper (where the iterated integrals are represented by their symbols).

When the letters are rational functions, the iterated integrals can be cast into MPLs. Supposing that $f_i(z) = 1/(z-z_i)$, the MPLs are defined as
\begin{equation}
\label{eq:MPL}
G(z_1,\,z_2,\ldots,z_n ;\, z) \equiv I(f_1,f_2,\ldots,f_n ;\, 0, z) \,.
\end{equation}
In the MPL cases, the ``trailing zero'' or ``tangential base point'' regularization implies that
\begin{equation}
    \label{eq:GPLtrailingzeros}
	G(\underbrace{0, \ldots, 0}_{r \text {-times }};\, z)=\frac{1}{r !} \ln ^{r}(z) \,.
\end{equation}
In our cases, the symbols of the solutions involve algebraic letters (square roots). They can be transformed into rational letters via variable changes. This is usually called ``rationalization''. Family (b) involves only one square root $\sqrt{1-4y}$, which can be rationalized by
\begin{equation}
	y = \frac{t}{(1+t)^2} \, , \quad t = \frac{1-\sqrt{1-4y}}{1+\sqrt{1-4y}} \, .
\end{equation}
Family (a) involves two square roots $\sqrt{1 \pm 4y}$. However, they never appear simultaneously in a symbol. Hence, we can rationalize them individually. For $\sqrt{1-4y}$, we use the same transformation as family (b), while for $\sqrt{1+4y}$, we use
\begin{equation}
	y = \frac{u}{(1-u)^2} \, , \quad u = \frac{\sqrt{1+4y}-1}{\sqrt{1+4y}+1} \, .
\end{equation}
After rationalization, the solutions can be easily written as MPLs of the new variables $t$ and $u$. We solve the sub-sector integrals in these two families up to transcendental weight 6 using the expressions of harmonic polylogarithms (HPL)~\cite{Remiddi:1999ew,Gehrmann:2000zt}, which consist of a subset of MPL. The analytic continuation beyond the vicinity of $y=0$ in sub-sectors can be performed just by the analytic continuation of HPL, which can be performed by packages like \cite{Maitre:2005uu}. The results of sub-sectors are given in the auxiliary files attached to this preprint.

In later sections, we will encounter iterated integrals with letters related to modular forms. In these cases, the base point is conventionally chosen as $\tau = i\infty$ where $\tau$ is the modular variable, and there is a factor of $2\pi i$ in each integration measure:
\begin{equation}
		I(\eta_1,\,\eta_2,\cdots, \eta_n; i\infty, \tau) = (2\pi i)^n\int_{i\infty}^\tau d \tau_1 \eta_1(\tau_1) \int_{i\infty}^{\tau_{1}}d \tau_2 \eta_2(\tau_2)\cdots \int_{i\infty}^{\tau_{n-1}} d \tau_{n} \eta_n(\tau_n) \,.
\end{equation}
In practice, one introduces the $q$ variable as $q = \exp(2\pi i \tau)$, and the iterated integrals become
\begin{equation}
		I(\eta_1,\,\eta_2,\cdots, \eta_n; 0, q) = \int_0^q\frac{d q_1}{q_1} \eta_1(q_1) \int_0^{q_{1}}\frac{d q_{2}}{q_{2}} \eta_2(q_2)\cdots \int_0^{q_{n-1}}\frac{d q_n}{q_n} \eta_n(q_n) \,.
\end{equation}
These iterated integrals can be evaluated via the $q$-expansion within the convergence region. In particular, the results for the top-sector integrals in terms of $q$-expansions are given in the auxiliary files attached to this paper.

It is worthwhile to remark on the concepts of ``weight'' that will be used throughout the paper. The first concept is ``transcendental weight''. As usual, we assign weight $-1$ to $\varepsilon$ and weight $n$ to the Riemann zeta value $\zeta_n$. For MPLs, the transcendental weight is equal to the depth of the iterated integral. For example, Eq.~\eqref{eq:MPL} has depth $n$ and weight $n$. It is a different story for iterated integrals involving elliptic kernels, as transcendental weights may not agree with the depth of the iterated integral \cite{Broedel:2018qkq}. As we will see, the boundary conditions in the limit $y \to 0$ are UT functions in the MPL sense. Combined with the $\varepsilon$-factorized differential equations for the top-sector integrals, our results are UT or pure functions in the elliptic sense \cite{Frellesvig:2023iwr}. Here, the ``elliptic transcendental weight'' is again equal to the depth of the iterated integral. Finally, we will also mention the concept of ``modular weight'', which is a property of modular forms under modular transformations.

\section{The top-sector integrals in family (a)}
\label{sect:familya}

We now turn to investigate the top sectors that involve elliptic curves. We begin with the simpler case of family (a), shown in the left panel of Fig.~\ref{fig:NPTri}. The differential equations of the two top-sector integrals $I_1$ and $I_2$ depend on each other and also depend on the sub-sector integrals. Our basic strategy is as follows: Firstly, we impose the maximal cut to eliminate the sub-sector integrals from the differential equations. We then use the Picard-Fuchs operator to obtain the $\varepsilon$-factorization within the top sector. Finally, we recover the sub-sector dependence and bring them to $\varepsilon$ by suitable subtractions (this step turns out to be unnecessary for family (a)). This strategy will also be applicable to family (b). With the $\varepsilon$-factorized differential equations at hand, we study the solutions for the top-sector integrals as functions of the modular variable and present numeric results.

\subsection{The Picard-Fuchs operator and $\varepsilon$-factorization}

The two top-sector integrals $I_1$ and $I_2$ in family (a) satisfy a system of two first-order differential equations, with dependence on sub-sector integrals $M_3,\ldots,M_{11}$. This system is equivalent to the following second-order differential equation for $I_1$:
\begin{align}
	L_{2}^{(\varepsilon)}(y) \, I_{1} &\equiv \sum_{n=0}^{2} r_{n}(y,\epsilon) \left( \frac{d}{dy} \right)^n I_{1} \nonumber
	\\
	&= \frac{1}{\varepsilon^2(16y-1)} \bigg[\overbrace{-8M_4+\frac{12}{\sqrt{1-4y}}M_5+28M_6-16M_7-10M_8}^{ f_{\rm sub}(y)}\bigg] \,,
	\label{eq:PFcross}
\end{align}
with the coefficients
\begin{equation}
    \label{eq:PFcoeffsa}
	\begin{aligned}
		r_{2}(y,\varepsilon) &= 1 \,, \\
		r_{1}(y,\varepsilon) &= \frac{-3+32y+2\varepsilon (-1+8y)}{y(1-16y)} \,, \\
		r_{0}(y,\varepsilon) &= \frac{4(1-9y+\varepsilon (1-6y))}{y^2(1-16y)} \,.
	\end{aligned}
\end{equation}
The superscript ``$(\varepsilon)$'' in the Picard-Fuchs operator $L_{2}^{(\varepsilon)}(y)$ reminds us that it depends on $\varepsilon$. The maximal cut for an integral of the form \eqref{eq:NPTriMomdef} amounts to taking the residue around zero for each of the propagator denominators $D_1, \ldots, D_6$. Under the maximal cut, the sub-sector integrals automatically vanish, and the above second-order differential equation for $I_1$ becomes homogeneous. The information about the top sector is then entirely contained in the operator $L_{2}^{(\varepsilon)}(y)$ that we are going to study. The basic idea towards constructing the $\varepsilon$ form is to observe that $L_{2}^{(0)}(y)$, which is obtained from $L_{2}^{(\varepsilon)}(y)$ by setting $\varepsilon=0$ and is closely related the elliptic curve, annihilates $I_1$ modulo $\mathcal{O}(\varepsilon)$ terms and sub-sector contributions. Then the solutions to $L_{2}^{(0)}(y)$ and a reformulation of $L_{2}^{(0)}(y)$ will hint us how to redefine a new master integral based on $I_1$ and the solutions of $L_{2}^{(0)}(y)$ in the following. 

The two independent kernels $\psi_0(y)$ and $\psi_1(y)$ of this operator, satisfying $L_{2}^{(0)}(y) \, \psi_k(y) = 0$, correspond to the two periods of the elliptic curve. We resort to the Frobenius method \cite{Frobenius+1873+214+235, inceODE:1956, agarwal2008ordinary} to solve for them in a suitable region of $y$. The Frobenius method can be used to construct series solutions to ordinary differential equations. For its applications in the context of Feynman integrals, we refer the readers to, e.g., \cite{Bonisch:2021yfw, Weinzierl:2022eaz}. In the following, we briefly outline the necessary steps. The first step is to study the regular singularities of the Picard-Fuchs operator. In our case, $L_2^{(0)}$ has three regular singular points: $\{0,\,1/16,\,\infty\}$. Here we work on the point $y=0$. The second step is then to derive the so-called indicial equation (characteristic equation) around this point. The indicial equation takes the form $P(\rho) = 0$, where $P(\rho)$ is a polynomial constructed from and of the same order as the Picard-Fuchs operator. The roots $\rho_i$ (called indicials or local exponents) determine the asymptotic behaviors of the solutions around the singular point. In our case, the indicial equation for the point $y=0$ reads $(\rho-2)^2=0$, which has a root with multiplicity 2: $\rho_0=\rho_1=2$. Here the indicial equation is said to have the maximal multiplicity, and the singular point is said to have the maximal unipotent monodromy (MUM). In this case\footnote{Around the MUM point, the local solutions present the maximal logarithmic pattern: the first solution $\psi_0$ is a power series, and the $(k+1)$-th solution $\psi_k$ has logarithms up to order $k$. More generally, when the roots, say $\rho_i$ and $\rho_j$, are separated by a non-zero integer, logarithmic structures may appear \textit{or not} \cite{Frobenius+1873+214+235, inceODE:1956, agarwal2008ordinary}. One needs to check specifically. }, the solutions will develop logarithmic asymptotic behaviors, and the logarithmic terms of the solution $\psi_k$ are completely fixed by $\{\psi_0,\ldots,\psi_{k-1}\}$ \cite{Frobenius+1873+214+235, inceODE:1956, agarwal2008ordinary}. For our $L_2^{(0)}$ around $y=0$, we have
\begin{equation}
    \label{eq:Frobeniusbasis}
	\begin{aligned}
		\psi_0(y) &= y^{\rho_0}\Sigma_0(y) \,, \quad \psi_1(y) = C_1 \, y^{\rho_0}\left[\Sigma_0(y)\ln y+\Sigma_1(y) \right] , \quad
	\end{aligned}
\end{equation}
where $\Sigma_{0}(y)$ and $\Sigma_{1}(y)$ are power series and can be determined iteratively. This basis is called the Frobenius basis. The above logarithmic structure can be intuitively understood as follows. We first assume that $\rho_1=\rho_0+\delta$ where $\delta$ is not an integer. According to the Frobenius method, there are two solutions with asymptotic behaviors $\psi_0 \sim y^{\rho_0}$ and $\psi_1 \sim y^{\rho_1}$. Now suppose we want to take the limit $\rho_1 \to \rho_0$, i.e., $\delta \to 0$. Directly taking $\delta \to 0$ in $\psi_1$ does not produce a solution independent of $\psi_0$. For a non-trivial limit, one needs to take a derivative with respect to $\delta$ before taking $\delta \to 0$. This leads to the logarithmic structure in \eqref{eq:Frobeniusbasis}. Similar structures are expected for higher-degree operators, as we will encounter in family (b).

According to the general form \eqref{eq:Frobeniusbasis}, we can write the local solutions as follows:
\begin{equation}
	\label{eq:periodsdef}
	\begin{aligned}
		\psi_{k}(y) = \frac{1}{(2 \pi i)^{k}} \sum_{j=0}^{k} \frac{\ln ^{j} y}{j !} \sum_{n=0}^{\infty} a_{k-j, n} \, y^{n+2} \,, \quad (k=0,1) \,.
	\end{aligned}
\end{equation}
The above choice fixes the relative normalization between the two solutions, and the overall normalization is chosen as $a_{0,0}=1$ for convenience. The relative normalization here leads to a Taylor-series relation between $q$ and $y$, see \eqref{eq:qya}. We solve the coefficients $a_{i,j}$ iteratively and obtain:
\begin{equation}
    \label{eq:Frobeniusa}
	\begin{aligned}
		\psi_{0}(y) &= y^2 \left( 1+4y+36y^2+400y^3+4900y^4 \right) + \mathcal{O}(y^7) \,, \\
		\psi_{1}(y) &= \frac{1}{2\pi i} \left[ \psi_{0}\ln y+y^2\left(8y+84y^2+\frac{2960}{3}y^3+\frac{37310}{3}y^4\right)+\mathcal{O}(y^7) \right] .
	\end{aligned}
\end{equation}
The modular variable is the ratio of the two periods: $\tau = \psi_1/\psi_0$. A function defined on the related elliptic curve can be Fourier-expanded in terms of the $q$ variable: $q \equiv \exp(2\pi i\tau)$. In particular, the $y$ variable can be treated as a function of $\tau$ or $q$ and vice versa. In the vicinity of $y=0$, the relations read
\begin{equation}
	\label{eq:qya}
	\begin{aligned}
		y(q) &= q-8q^2+44q^3-192q^4+718q^5-2400q^6+\mathcal{O}(q^7) \,, \\
		q(y) &= y+8y^2+84y^3+992y^4+12514y^5+164688y^6+\mathcal{O}(y^7) \,.
	\end{aligned}
\end{equation}
Note that the first coefficients in the above series are fixed to unity by the relative normalization between $\psi_0$ and $\psi_1$ chosen in Eq.~\eqref{eq:periodsdef}. The $q$-expansion of $y$ in Eq.~\eqref{eq:qya} can be expressed in terms of Dedekind eta-quotients, and we find that
\begin{equation}
    \label{eq:ytaua}
	y(\tau) = \frac{\eta^8(\tau)\eta^{16}(4\tau)}{\eta^{24}(2\tau)}=\frac{1}{16}\lambda(2\tau) \, ,
\end{equation}
where $\eta(\tau)$ is the Dedekind eta function and $\lambda(\tau)$ is the modular $\lambda$ function. The above series can also be found in the Online Encyclopedia of Integer Sequences (OEIS) \cite{OEIS:A005798}. This is a well-defined extension of the series-expansion relation in Eq.~\eqref{eq:qya}, i.e., given a $\tau$, one obtains a $y$. However, it is not straightforward to write down the inverse map from $y$ to $\tau$ or $q$. We will show in the next subsection how to perform the analytic continuation for the inverse map. After that, given a $y\in\mathbb{R}+i0$, we obtain a $\tau$ via $\tau(y)$ and return the same value of $y$ via $y\big(\tau(y)\big)$ without ambiguity. The rest of this subsection does not rely on this continuation. We only require the modular variable to be defined as the ratio of two periods controlled by the Picard-Fuchs operator and formulate the following as generally as possible.

Any function or operator $f$ of $y$ can be treated as that of $\tau$ or $q$. We will abuse the notation a bit and write $f(y)$, $f(\tau)$, or $f(q)$ depending on the context. For differential operators, the Jacobian is given by
\begin{equation}
	\label{eq:Jacobian}
	J(y) = \frac{1}{2\pi i} \frac{d y}{d\tau} = \Theta_q\,y \,, \quad \text{with} \quad \Theta_q \equiv q \frac{d}{d q} = J(y) \frac{d}{dy} \,.
\end{equation}
It can be expressed in terms of the Wronskian
\begin{equation}
	W(y)\equiv 2\pi i\left[\psi_{0}\frac{d}{dy}\psi_{1}-\psi_{1}\frac{d}{dy}\psi_{0}\right]=\frac{y^3}{1-16y} \, ,
\end{equation}
as
\begin{equation}
	\label{eq:JacobianWronskian}
	J(y) = \frac{\psi_{0}^{2}}{W(y)} \, .
\end{equation}
The above two identities can be derived from the differential equation satisfied by $\psi_{k}$ and the definition for $\tau=\psi_1/\psi_0$.
Note that the initial Picard-Fuchs operator $L_2^{(0)}(y)$ is an irreducible second-order operator. However, with the variable change, one can verify that it factorizes in $q$-space as:
\begin{equation}
	\label{eq:PFfacta}
	L_2^{(0)}(y) = \frac{\psi_0}{J^2} \, \Theta_q^2 \, \frac{1}{\psi_0} \,.
\end{equation}
This is a special case of the generalized factorization of Calabi-Yau operators~\cite{Pogel:2022vat, Duhr:2022dxb, Bogner:13}. After the analytic continuation of the modular map, Eq.~\eqref{eq:PFfacta} holds for all values of $y$. The factorization pattern is the starting point of the canonical basis in the top sector below.

We now proceed to choose the ansatz for the top-sector canonical MIs as
\begin{equation}
    \label{eq:ansatza}
	\begin{aligned}
		M_{1} &= \varepsilon^4\frac{I_{1}}{\psi_{0}} \,, \\
		M_{2} &= \frac{J(y)}{\varepsilon}\frac{d}{dy}M_{1}-F_{11}(y)\,M_{1} \,,
	\end{aligned}
\end{equation}
We require the two MIs to satisfy the following $\varepsilon$-factorized differential equations under the maximal cut:
\begin{equation}
    \label{eq:epstopa}
	J(y)\frac{d}{d y}\left(\begin{array}{c} M_{1} \\ M_{2} \end{array} \right)^{\rm mc}= \varepsilon\, A^{\rm mc}(y) \left(\begin{array}{c} M_{1} \\ M_{2} \end{array} \right)^{\rm mc} \, .
\end{equation}
These lead to constraints on the coefficient function $F_{11}$, which can be easily solved as
\begin{equation}
	F_{11}(y) = \frac{1-8y}{y^4}\psi_{0}^{2} \,.
\end{equation}
The connection matrix for the top sector takes the following form
\begin{equation}
	A^{\rm mc}=\left(
		\begin{array}{cc}
			F_{11} & 1 \\ F_{11}^2 & F_{11}
		\end{array}
	\right).
\end{equation}

We now bring back the sub-sector dependence with the inhomogeneous term in \eqref{eq:PFcross}. Note that the function $f_{\text{sub}}(y)$ consists of UT MIs and the $\varepsilon$-dependence is already factorized out in Eq.~\eqref{eq:PFcross}. These tell us that the differential equations for $M_1$ and $M_2$ are automatically $\varepsilon$-factorized without the maximal cut:
\begin{equation}
    \label{eq:epstopa}
	J(y)\frac{d}{d y}\left(\begin{array}{c} M_{1} \\ M_{2} \end{array} \right)= \varepsilon\, A^{\rm mc}(y) \left(\begin{array}{c} M_{1} \\ M_{2} \end{array} \right) + \varepsilon\frac{16y-1}{y^6}\psi_0^3(y)f_{\rm sub}(y)\,\begin{pmatrix}
		0\\
		1
	\end{pmatrix} \, .
\end{equation}
Hence, we have obtained the $\varepsilon$-form differential equations for family (a), with the canonical MIs in Eqs.~\eqref{eq:ansatza} and \eqref{eq:subUTa}.

\subsection{Elliptic curve and analytic continuation of the modular map}

The map between $y$ and $q$ in Eqs.~\eqref{eq:Frobeniusa} and \eqref{eq:qya} only works in the vicinity of $y=0$. In this subsection, we analytically continue this map to be valid for the whole kinematic space. This is achieved by studying the associated elliptic curve. We formulate the $\varepsilon$ form in the top sector in terms of $y$. Hence all the formulae in the previous section carry on in terms of $\tau$ with the continued map, valid for all kinematical values (except the threshold point).

\begin{figure}[t!]
  \centering
  \includegraphics[width=0.4\textwidth]{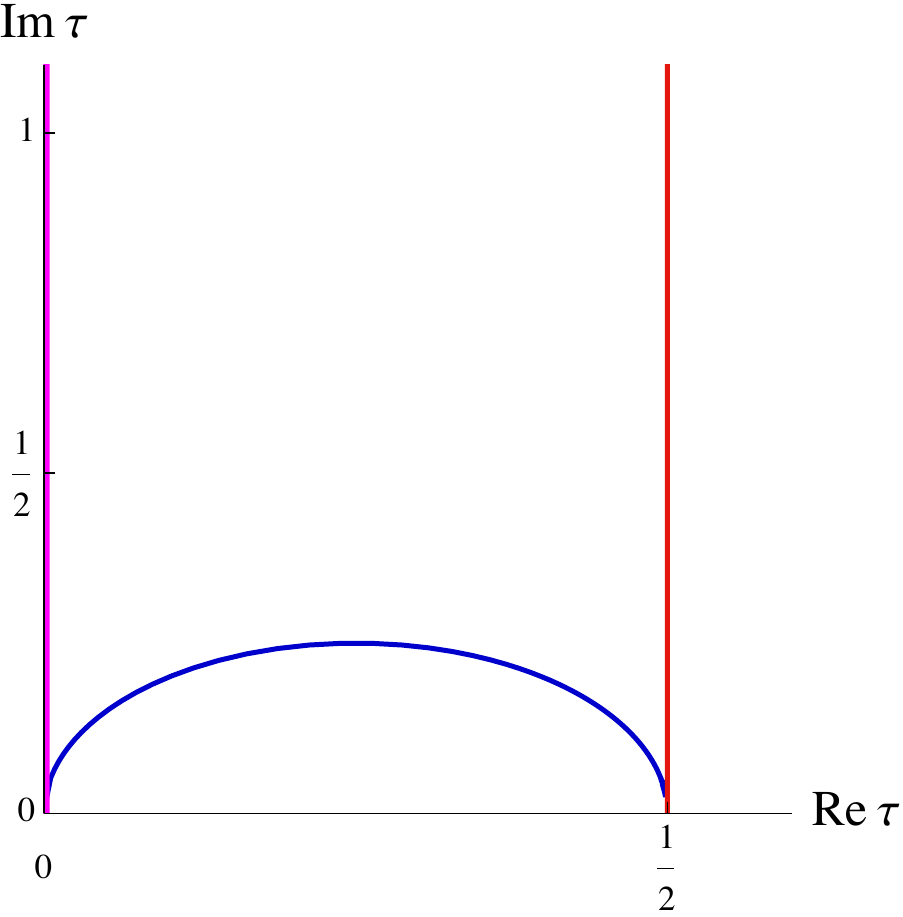}
  \hspace{1cm}
  \includegraphics[width=0.4\textwidth]{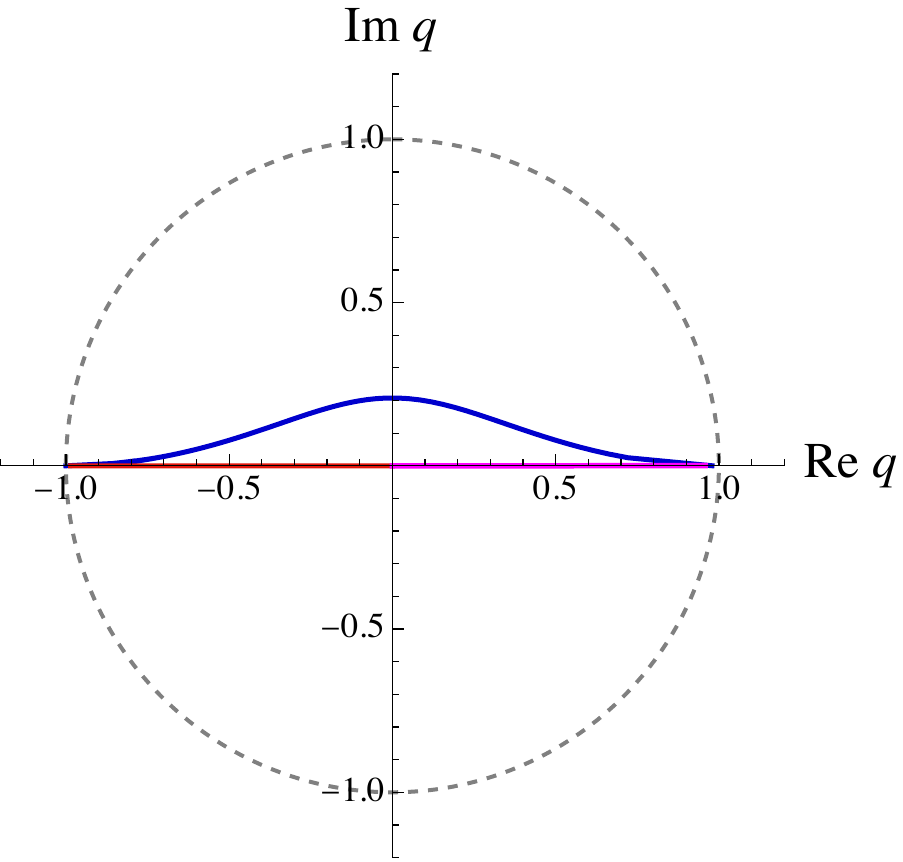}
  \caption{\label{fig:patha}The behaviors of $\tau$ and $q$ as functions of $y$. The red path corresponds to $y\in (-\infty, 0)$, the magenta one corresponds to $y\in [0,\,1/16)$, and the blue one corresponds to $y\in [1/16,+\infty) $. }
\end{figure}

\begin{figure}[t!]
  \centering
  \includegraphics[width=10cm]{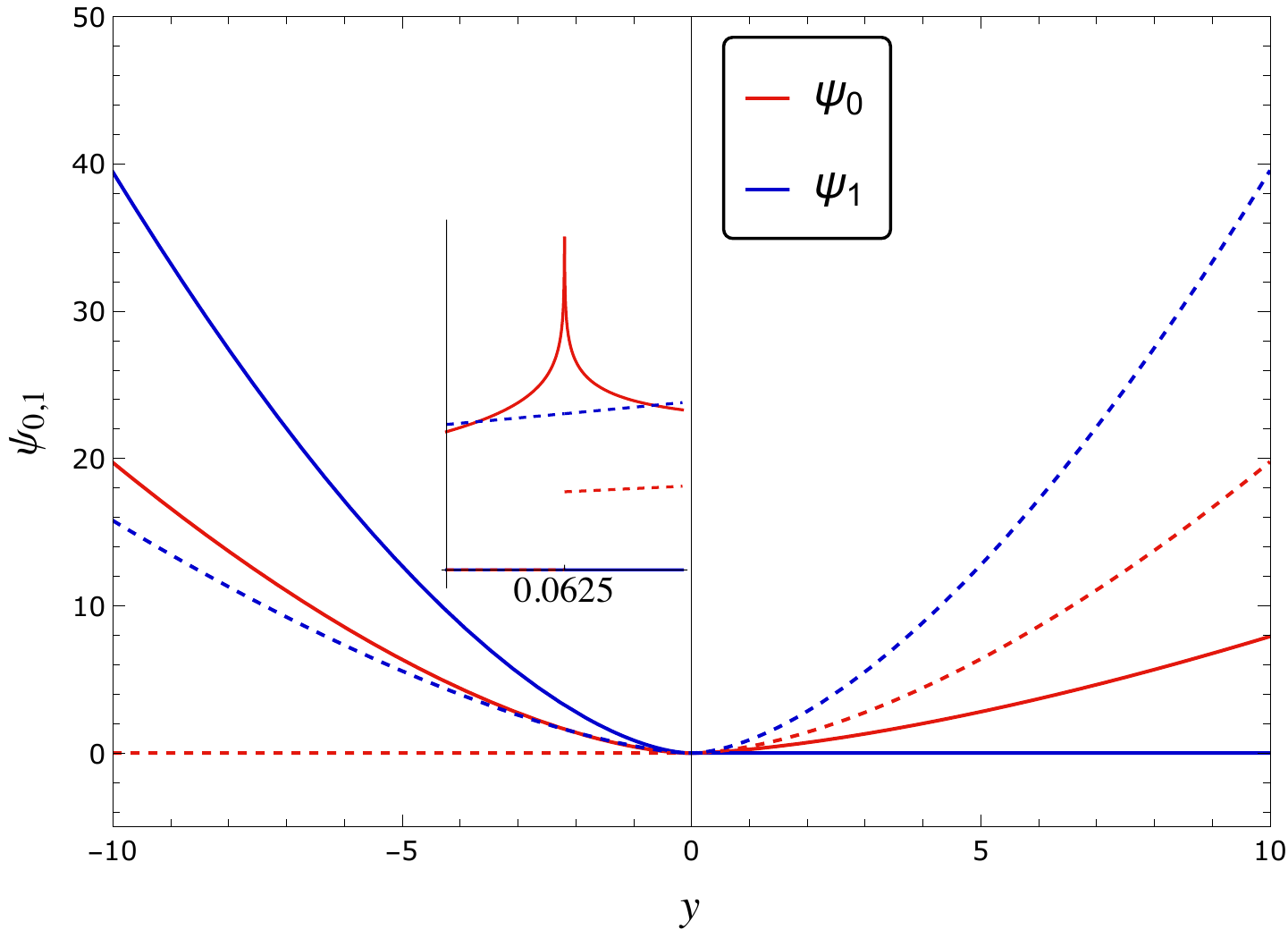}
  \caption{\label{fig:periodsa}The two periods as functions of $y$. The solid lines represent real parts, and the dashed lines are imaginary parts. There are three cusps for $\Gamma_1(4)$: $\tau=0,1/2,i\infty$. With the normalization in Eq.~\eqref{eq:anacperiodsa}, $\psi_0$ is holomorphic around $\tau=i\infty$ ($y=0$). However, it is not continuous around the cusp $\tau=1/2$ ($y=1/16$), shown as a zoomed sub-plot in the figure.}
\end{figure}

Taking the maximal cut of $I_1$ in the Baikov representation, we obtain the associated elliptic curve:
\begin{equation}
    \label{eq:Ea}
	E: \; v^2 = \left(u-u_1\right)\left(u-u_2\right)\left(u-u_3\right)\left(u-u_4\right),
\end{equation}
where the four roots are
\begin{equation}
    \label{eq:roota}
	u_{1}=0 \,, \quad u_{2}=\frac{1-\sqrt{1-16y}}{2} \,, \quad u_{3}=\frac{1+\sqrt{1-16y}}{2} \,, \quad u_{4}=1 \,.
\end{equation}
In the region $0 < y < 1/16$, the four roots are ordered as $u_1 < u_2 < u_3 < u_4$. We define the elliptic modulus as
\begin{equation}
    \label{eq:modulusa}
	k^2=\frac{(u_2-u_1)(u_4-u_3)}{(u_3-u_1)(u_4-u_2)}=\left(\frac{1-\sqrt{1-16y}}{1+\sqrt{1-16y}}\right)^2 .
\end{equation}
The periods of the elliptic curve reside in the solution space of the Picard-Fuchs operator $L_2^{(0)}$, which is spanned by the two complete elliptic integrals of the first kind, $\{K(k^2),\,iK(1-k^2)\}$. It attempts to relate the modular variable $\tau$ to the ratio of these two elliptic integrals. However, as $k^2$ changes with $y$, the complete elliptic integrals will develop a discontinuity when $y$ crosses a branch cut. As a result, the periods naively defined by the complete elliptic integrals as functions of $\tau$ are not smooth. This can be fixed by compensating for the discontinuity when crossing the branch cuts. In this context, we define the two periods $\bar{\psi}_0(y)$ and $\bar{\psi}_1(y)$ as linear combinations of the two elliptic integrals:
\begin{equation}
	\label{eq:anacperiodsa}
	\begin{aligned}
		\begin{pmatrix}
			\bar{\psi}_1(y)\\
			\bar{\psi}_0(y)
		\end{pmatrix}=
		\frac{2}{\pi}\frac{y^2}{\sqrt{(u_3-u_1)(u_4-u_2)}} \, \gamma(y) \begin{pmatrix}
			i\,K(1-k^2)\\
			K(k^2)
		\end{pmatrix},
	\end{aligned}
\end{equation}
where the monodromy matrix $\gamma(y)$ is given by
\begin{equation}
	\label{eq:monodromy}
		\gamma(y) =
		\begin{cases}
		\begin{aligned}
		&\begin{pmatrix}
			1 & 2\\
			0 & 1
		\end{pmatrix},\quad y < 0 \text{ or } y \geq  \frac{1}{8} \, ,
		\\
		&\begin{pmatrix}
			1 & 0\\
			0 & 1
		\end{pmatrix},\quad 0 \leq y < \frac{1}{8} \,.
		\end{aligned}
		\end{cases}
\end{equation}
The matrix is determined by the discontinuity~\cite{Bogner:2017vim} of the complete elliptic integral $K(k^2)$, whose branch cut in terms of $k^2$ is $[1,\infty)$. At $y=1/8+i0$, $1-k^2$ goes across the branch cut of $K(1-k^2)$, and one needs to compensate the change of the imaginary part to make the function continuous. Performing the series expansion around $y = 0$ and comparing with Eq.~\eqref{eq:Frobeniusa}, we may identify $\psi_0 = \bar{\psi}_0$ and $\psi_1 = \bar{\psi}_1/4$, and
\begin{equation}
   \label{eq:tauya}
   \tau = \frac{\bar{\psi}_1(y)}{4\bar{\psi}_0(y)} \, .
\end{equation}
It is easy to verify that the above function is indeed the inverse map of Eq.~\eqref{eq:ytaua} in the entire phase space. Thus Eq.~\ref{eq:ytaua} and Eq.~\ref{eq:tauya} give one-to-one maps between $y$ and $\tau$. Interestingly, we find that the three kinematic singular points are exactly mapped to the three cusps of the congruence subgroup $\Gamma_1(4)$:
\begin{equation}
   \label{eq:cuspsa}	
   \tau(y=0) = i\infty\,,\quad \tau(y=\infty) = \frac{1}{2}\,,\quad \tau(y=1/16) = 0 \,.
\end{equation}
We show the behaviors of $\tau$ and $q$ as functions of $y$ in Fig.~\ref{fig:patha}, and the two periods are shown in Fig.~\ref{fig:periodsa}.

\subsection{Letters and modular forms}

We may now consider letters in the $\varepsilon$-factorized differential equations as functions of the modular variable $\tau$. It is instructive to study the properties of these functions under the modular transformation
\begin{equation}
\tau \to \frac{a \tau + b}{c \tau + d} \, ,
\end{equation}
with $a, b, c, d\in\mathbb{Z}$ and $ad-bc=1$. As mentioned earlier, the relevant congruence subgroup is $\Gamma_1(4)$, meaning that $a,d \equiv 1 \pmod 4$ and $c \equiv 0 \pmod 4$. The functions should have specific scaling behaviors under this group transformation.

First of all, the kinematic variable $y$ as a function of $\tau$ should not change under the transformation, i.e.,
\begin{equation}
	y\left(\frac{a\tau+b}{c\tau + d}\right) = y(\tau) \,.
\end{equation}
In other words, $y(\tau)$ is a modular function under the congruence subgroup. It is a meromorphic function in the complex upper half-plane $\mathbb{H}$ and at the cusps. In Eq.~\eqref{eq:ytaua}, we have given the analytic expression of $y(\tau)$ in terms of an eta quotient, which is valid in the whole domain.

The periods and the coefficients in the differential equations are, in general, modular forms, transforming with certain modular weight $k$:
\begin{equation}
	f\left(\frac{a\tau+b}{c\tau + d}\right) = (c\tau+d)^{k}f(\tau) \,,
\end{equation}
which are holomorphic in $\mathbb{H}\cup\{\rm cusps\}$.
The building block at hand is $\psi_0$, a modular form with weight 1. Note that the dimension of the modular weight-1 Eisenstein subspace of $\Gamma_1(4)$ is 1, and that of the modular weight-2 Eisenstein subspace is 2. They are spanned by $\{e_{1,1}\}$ and $\{e_{1,2},\,e_{2,2}\}$ respectively:
\begin{equation}
	\label{eq:psi0eisenstrin}
	\psi_0 = \frac{\lambda(2\tau)^2}{64}\,e_{1,1} \,.
\end{equation}
We refer the readers to \cite{Weinzierl:2022eaz} for their explicit definitions. The Jacobian $J$ and the coefficient $F_{11}$ has modular weight 2, while $F_{11}^2$ is of modular weight 4. The mixing with sub-sectors involves a modular form of weight 3. It is convenient to introduce another basis for the weight 1 and weight 2 modular forms:
\begin{equation}
	\label{eq:modularbasisbnew}
	b_{1,1} = \frac{\psi_0}{y^2}\,,\quad b_{1,2} = \frac{1-8y}{y^4}\psi_0^2\,,\quad b_{2,2} = \frac{1-16y}{y^4}\psi_0^2\,,
\end{equation}
which is related to the Eisenstein basis by
\begin{equation}
	\label{eq:modularbasisbnewcont}
	e_{1,1}=\frac{1}{4}b_{1,1}\,,\quad e_{1,2} = \frac{1}{6}b_{1,2}-\frac{1}{8}b_{2,2}\,,\quad e_{2,2} = \frac{1}{4}b_{1,2}-\frac{1}{8}b_{2,2}\,.
\end{equation}

Four letters in the differential equation are modular forms. They are
\begin{equation}
	\label{eq:modularlettersa}
	\begin{aligned}
		\eta_{1,2} = b_{1,2}\,,\quad \eta_{2,2} = b_{2,2}\,,\quad \eta_3 = b_{1,1}b_{2,2}\,,\quad \eta_4 &= b_{1,2}^2\,.
	\end{aligned}
\end{equation}
Five extra letters are not modular forms, given by
\begin{equation}
	\label{eq:eq:notmodularlettersa}
	\begin{aligned}
		\varrho = \frac{\eta_3}{\sqrt{1-4y}}\,,\quad \vartheta = \frac{\eta_{2,2}}{\sqrt{1-4y}}\,,\quad \varphi = \frac{\eta_{2,2}}{\sqrt{1+4y}}\,,\quad \varpi_1 = \frac{\eta_{2,2}}{1-4y}\,,\quad \varpi_2 = \frac{\eta_{2,2}}{1+4y} \, .
	\end{aligned}
\end{equation}
The differential equation for the whole family (a) can then be reformulated as
\begin{equation}
	\label{eq:epsformafinal}
	\begin{aligned}
			\frac{1}{2\pi i}\frac{d \vec{M}}{d\tau} =
		\varepsilon\begin{pmatrix}
		\eta_{1,2} & 1 & 0 & 0 & 0 & 0 & 0 & 0 & 0 & 0 & 0\\
		\eta_4 & \eta_{1,2} & 0 & 8 \eta _3 & -12 {\color{red}\varrho} & -28 \eta _3 & 16 \eta _3 & 10 \eta _3 & 0 & 0 & 0\\
		0 & 0 & 0 & 0 & 0 & 0 & -\eta_{2,2} & 0 & 0 & 0 & 0 \\
		0 & 0 & 0 & \eta_{2,2} & 0 & -\eta_{2,2} & \frac{\eta_{2,2}}{2} & 0 & 0 & 0 & 0 \\
		0 & 0 & 0 & 0 & \eta_{2,2}+2{\color{red}\varpi_1} & 3{\color{red}\vartheta } & 0 & 0 & 0 & -{\color{red}\vartheta} & -{\color{red}\vartheta} \\
		0 & 0 & 0 & 0 & -{\color{red}\vartheta} & -\eta_{2,2} & 0 & 0 & 0 & 0 & 0 \\
		0 & 0 & 0 & 0 & 0 & 0 & 0 & -\eta_{2,2} & 0 & 0 & 0 \\
		0 & 0 & 0 & 0 & 0 & 0 & 0 & -\eta_{2,2} & -2{\color{red}\varphi} & 0 & 0 \\
		0 & 0 & 0 & 0 & 0 & 0 & 0 & 3{\color{red}\varphi} & \eta_{2,2}+3{\color{red}\varpi_2} & 0 & -{\color{red}\varphi} \\
		0 & 0 & 0 & 0 & 0 & 0 & 0 & 0 & 0 & \eta_{2,2} & 0 \\
		0 & 0 & 0 & 0 & 0 & 0 & 0 & 0 & 0 & 0 & 0
		\end{pmatrix}
		\vec{M},
	\end{aligned}
\end{equation}
with $\vec{M}=(M_1, M_2, M_3, \cdots, M_{11})^T$. And we have colored the extra letters for convenience, as for family (b) later on. 

Comparing the above equation with Eq.~\eqref{eq:epssuba}, we see that the letters with modular weight 2 actually correspond to algebraic letters in terms of $y$. Here we note that $\eta_{1,2}$ and $\eta_{2,2}$ are weight-2 modular forms, and $y$ is a weight-0 modular function of $\tau$. The derivative with respect to $\tau$ increases the modular weight by 2. An example is the Jacobian $J(y(\tau))$, which has weight 2. The relations between the $\tau$-representation and the $y$-representation of these weight-2 letters are given by:
\begin{equation}
	\label{eq:dloglettersa}
	\begin{aligned}
		\eta_{1,2}\cdot2\pi i\,d\tau &= d\log y-\frac{1}{2}d\log (1-16y)\,,\\
		\eta_{2,2}\cdot2\pi i\,d\tau &= d\log y\,,\\
		\vartheta\cdot2\pi i\,d\tau &= d\log\frac{1-\sqrt{1-4y}}{1+\sqrt{1-4y}} = d\log t\,,\\
		\varphi\cdot2\pi i\,d\tau &= d\log\frac{\sqrt{1+4y}-1}{\sqrt{1+4y}+1} = d\log u\,,\\
		\varpi_1\cdot2\pi i\,d\tau &= d\log y-d\log(1-4y)\,,\\
		\varpi_2\cdot2\pi i\,d\tau &= d\log y-d\log(1+4y)\,.
	\end{aligned}
\end{equation}
The weight-3 and weight-4 letters ($\eta_4$, $\eta_3$ and $\varrho$), on the other hand, are genuinely elliptic and do not have $\mathrm{d}\log$ representations. It is worth mentioning that modular weight 2 does not guarantee $\mathrm{d}\log$ representations, either. For example, for three-loop Banana integrals, whose geometric object is a K3 surface, logarithmic differential one-forms are insufficient for modular weight-2 letters. See \cite{Pogel:2022vat} for an extensive discussion.

\subsection{Boundary conditions}
\label{sect:boundarya}

We need the boundary conditions to solve the differential equations, which we choose as $y = 0$. Since this is a singular point, we need to calculate the asymptotic behaviors of the canonical MIs in the limit $y \to 0$. The sub-sector integrals are simple, and we collect the boundary conditions in the Appendix. For the top sector, it is enough to calculate $I_1$.

From the indicial equation of $L_2^{(0)}$ around $y=0$, we know that $I_1$ scales as $y^2$ for small $y$, with some logarithmic corrections. Then the boundary terms are given by these logarithmic corrections. They can be calculated by Mellin-Barnes techniques. The Mellin-Barnes representation of $I_1$ reads:
\begin{equation}
	\label{eq:MBa}
	\begin{aligned}
		I_1 = e^{2\varepsilon\,\gamma_E} \int \prod_{i=1}^3\frac{d z_i}{2\pi i}\, y^{2+2\varepsilon+z_1} \Gamma (-z_1) \Gamma (-z_2)  \Gamma (-z_3)\frac{\Gamma (1+z_2)^2\Gamma (-\varepsilon -z_1)^2\Gamma (1+z_2+z_3)}{\Gamma (-2 \epsilon -2 z_1) \Gamma (-3 \epsilon -z_1)}\\
		\times \frac{ \Gamma (-1-2 \varepsilon -z_1-z_2) \Gamma (1+2 \epsilon +z_1-z_3) \Gamma (-2 \epsilon -z_1+z_3)^2 \Gamma (1-\epsilon +z_2+z_3)}{ \Gamma (-2 \epsilon -z_1+z_2+z_3+1)^2} \,,
	\end{aligned}
\end{equation}
and then we perform the asymptotic expansion for $y \to 0$. With the help of \texttt{MBTools} \cite{Belitsky:2022gba} and \texttt{XSummer} \cite{Moch:2005uc}, we obtain the first three orders in $\varepsilon$:
\begin{equation}
    \label{eq:boundaryI1a}
	\begin{aligned}
		\frac{I_{1}}{y^2}\bigg|_{y\to 0} =&\, \frac{1 }{3}L_y^4-6\zeta_2  L_y^2-40 \zeta_3 L_y-49\zeta_4+\varepsilon\bigg[\frac{1}{5}L_y^5-\frac{20\zeta_2}{3}L_y^3-42\zeta_3 L_y^2-29\zeta_4L_y\\
		& + 64\zeta_2\zeta_3+32\zeta_5\bigg]+\epsilon^2\bigg[\frac{7}{90}L_y^6-\frac{11\zeta_2}{3}L_y^4-32\zeta_3L_y^3-69\zeta_4L_y^2 \\
		&  +\left(124\zeta_2\zeta_3-236\zeta_5\right)L_y-280\zeta_6+246\zeta_3^2\bigg]+\mathcal{O}(\varepsilon^3), \,
	\end{aligned}
\end{equation}
where $L_y=\log y$.

It is straightforward to derive the boundary conditions for $M_1$ and $M_2$ from that of $I_1$. Using $\psi_0=y^2+\mathcal{O}(y^3)$, $F_{11}=1+\mathcal{O}(y)$, $J(y)=y+\mathcal{O}(y^2)$ and $y = q + \mathcal{O}(q^2)$, we have:
\begin{equation}
	\label{eq:familyaBD}
	\begin{aligned}
		M_1\big|_{y\to 0} &= \varepsilon^4 \left[\frac{1}{3}L_q^4-6\zeta_2 L_q^2-40\zeta_3 L_q-49\zeta_4\right]+\varepsilon^5\bigg[\frac{1}{5}L_q^5-\frac{20}{3} \zeta _2 L_q^3-42 \zeta _3 L_q^2\\
		&\hspace{5cm}-29 \zeta _4 L_q+64 \zeta _2 \zeta _3+32 \zeta _5\bigg]+\mathcal{O}(\varepsilon^6),\\
		M_2\big|_{y\to 0} &= \varepsilon^3\left[\frac{4 }{3}L_q^3-12 \zeta _2 L_q-40 \zeta _3\right]+\varepsilon^4\left[\frac{2 }{3}L_q^4-14 \zeta _2 L_q^2-44 \zeta _3 L_q+20 \zeta _4\right]+\mathcal{O}(\varepsilon^5),
	\end{aligned}
\end{equation}
where $L_q=\log q$.

\subsection{Results}
\label{scet:resultsa}

\begin{figure}[t!]
	\centering
	\includegraphics[width=7cm]{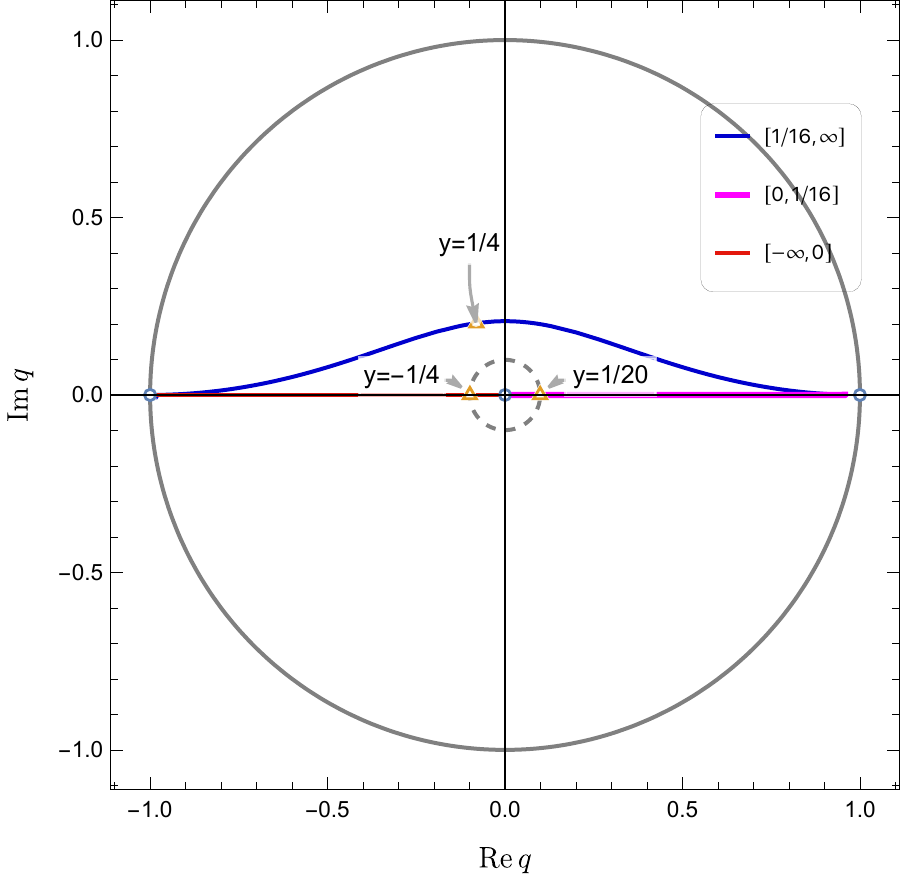}
	\caption{\label{fig:convergenta}Radius of convergence in $q$-space for the series expansion around $q=0$ for family (a), and the corresponding values of $y$.}
\end{figure}

The solutions to the differential equations can be formally written as iterated integrals over the letters, with the boundary conditions obtained above.

Armed with all the above, it is straightforward to write down the leading terms of $M_1$ and $M_2$ in $\varepsilon$ expansion here:
\begin{equation}
	\label{eq:familyatopana}
	\begin{aligned}
		M_2 &= \varepsilon^3\big[-40\zeta_3-12\zeta_2 \, I(\varrho;i\infty, \tau)+20\,I(\eta_3,\varphi,\varphi; i\infty, \tau)-12\,I(\varrho, \vartheta, \eta_{2,2}; i\infty, \tau)\big]+\mathcal{O}(\varepsilon^4),\,\\
		M_1 &= \varepsilon^4\big[ -49\zeta_4 -40\zeta_3 L_q-12\zeta_2 I(1, \rho; i\infty, \tau) +20\,I(1, \eta_3,\varphi,\varphi; i\infty, \tau)\\
		&\hspace{5.4cm}-12\,I(1, \varrho, \vartheta, \eta_{2,2}; i\infty, \tau)\big]+\mathcal{O}(\varepsilon^5).
	\end{aligned}
\end{equation}

Within the radius of convergence, the iterated integrals can be efficiently computed via $q$-expansion.\footnote{There are alternative ways to evaluate these integrals. See, e.g., the algorithm \cite{Walden:2020odh} implemented in \texttt{GiNaC} \cite{Bauer:2000cp}, and Ref.~\cite{Duhr:2019rrs}.} The convergence radius of the expansion is determined by the nearest singularity in relevant letters. We emphasize that the modular form letters are holomorphic in the upper half plane $\mathbb{H}$. Thus the singularities are present only in the algebraic letters from sub-sector dependence. Three singular points\footnote{The point $y=1/4$ is only relevant for the sub-sector integral $M_5$.} affect the top-sector integrals: $y = 0, \, -1/4, \, \infty$. To cover the whole phase space, we can perform the expansion around each of these points. Here we only consider the expansion around $y = 0$ (i.e., $q = 0$). We show the radius of convergence for this expansion in Fig.~\ref{fig:convergenta}, which is constrained by the singularity at $y=-1/4$. The dashed circle in the figure represents the convergence region. For real values of $y$, this region corresponds to $y \in (-1/4,1/20)$, i.e., $x = s/m^2 \in (-\infty, -20) \cup (4,\infty)$.

\begin{figure}[t!]
	\centering
	\includegraphics[width=0.45\textwidth]{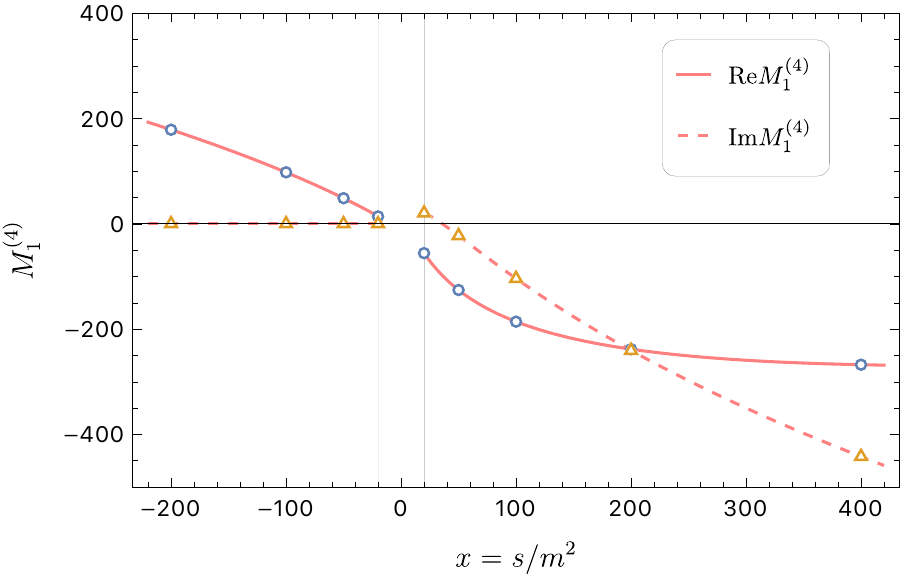}
	\hspace{0.5cm}
	\includegraphics[width=0.45\textwidth]{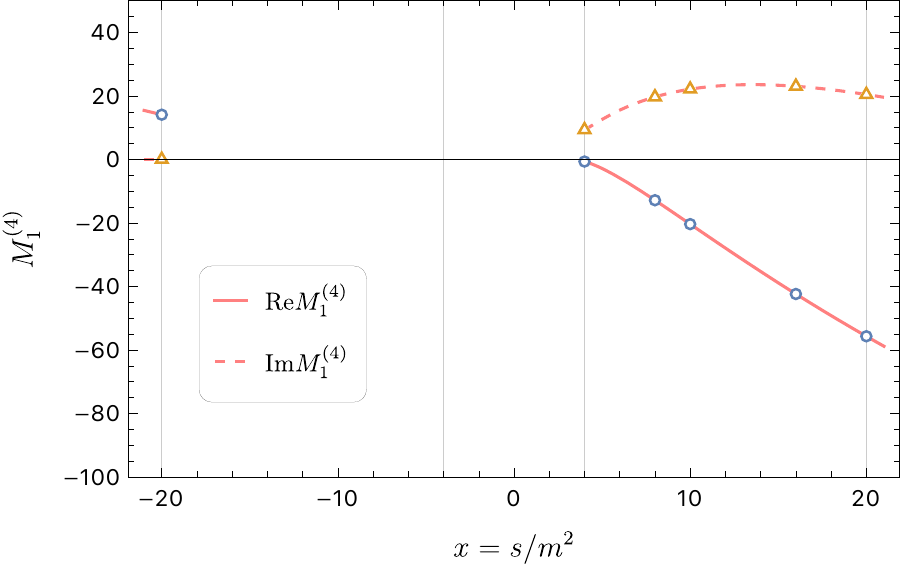}
	\\
	\includegraphics[width=0.45\textwidth]{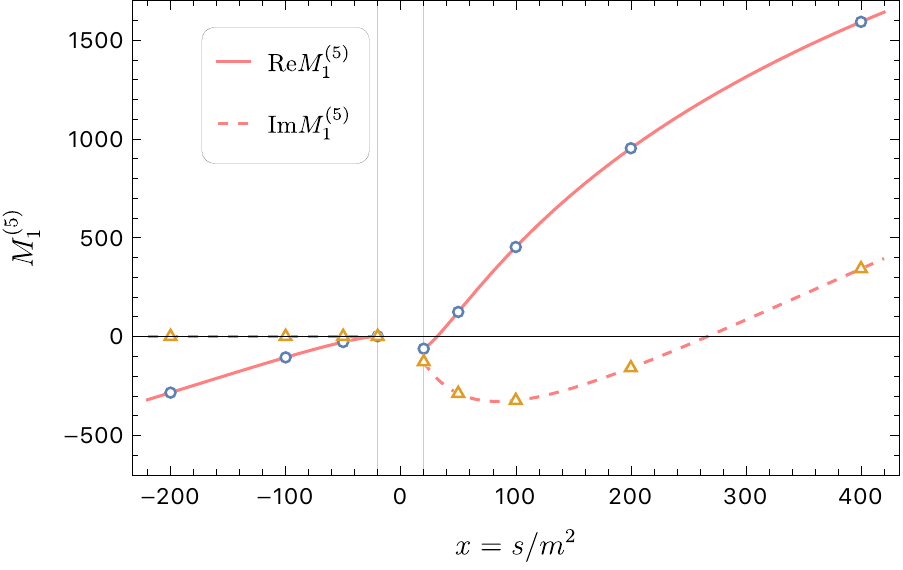}
	\hspace{0.5cm}
	\includegraphics[width=0.45\textwidth]{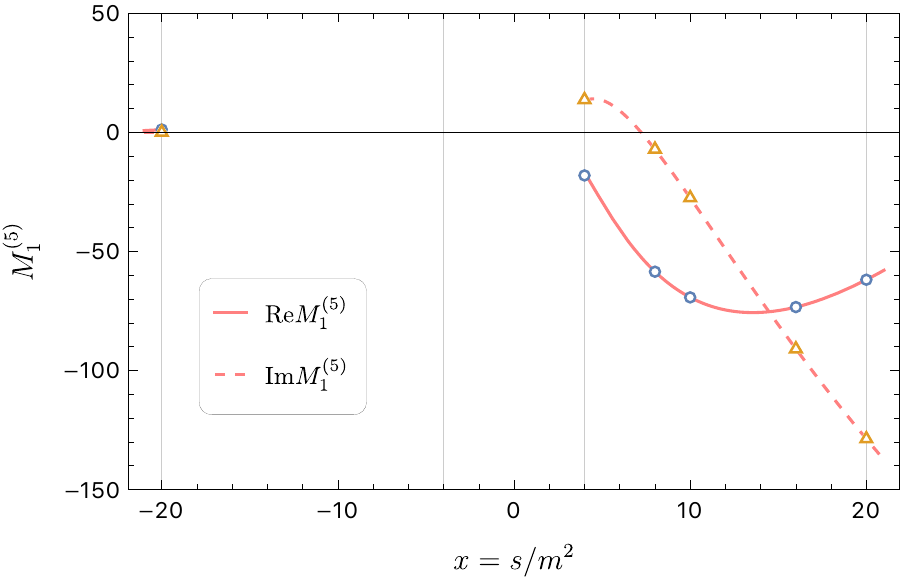}
	\\
	\includegraphics[width=0.45\textwidth]{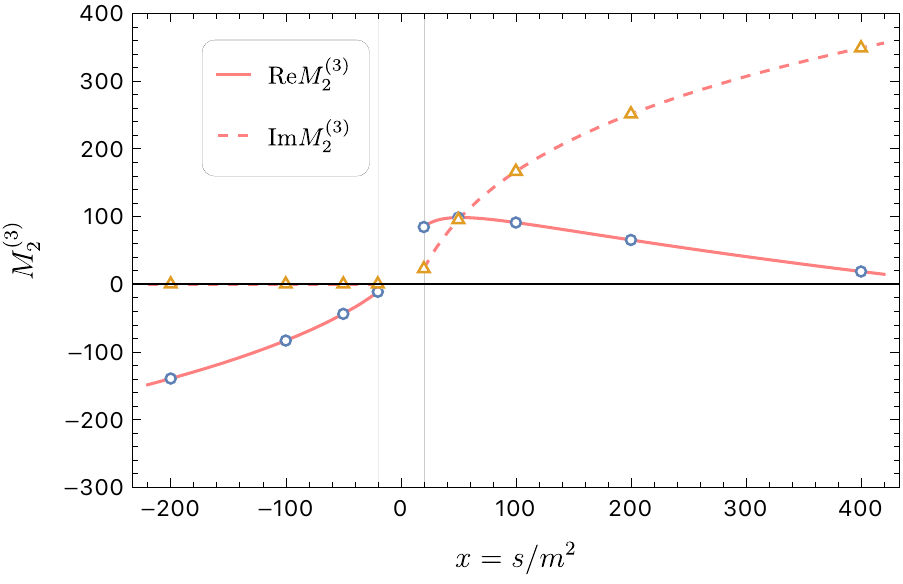}
	\hspace{0.5cm}
	\includegraphics[width=0.45\textwidth]{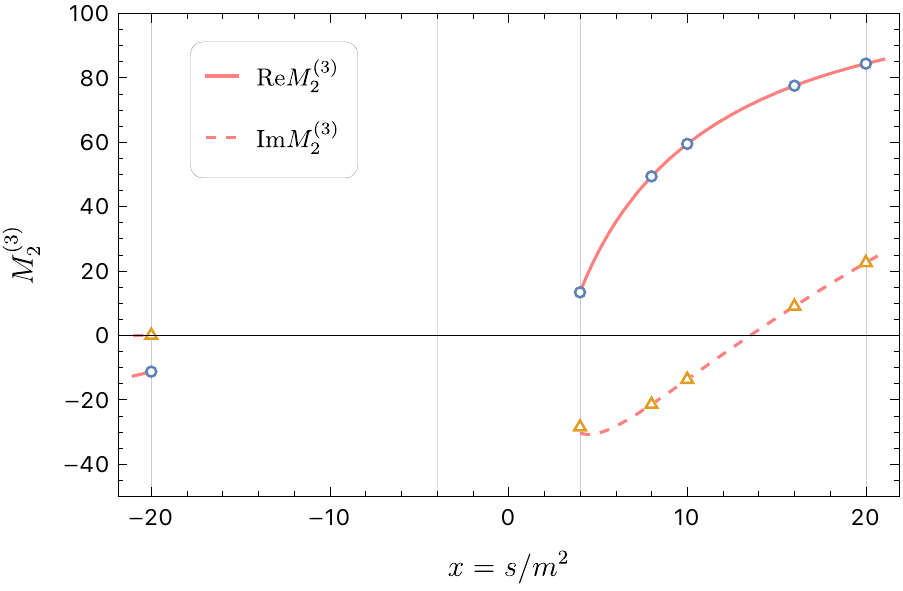}
	\\
	\includegraphics[width=0.45\textwidth]{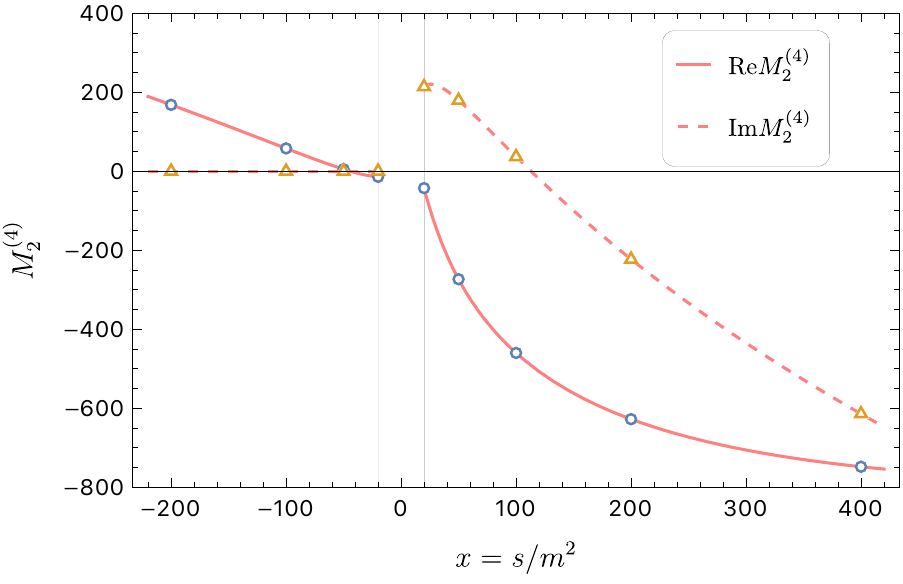}
	\hspace{0.5cm}
	\includegraphics[width=0.45\textwidth]{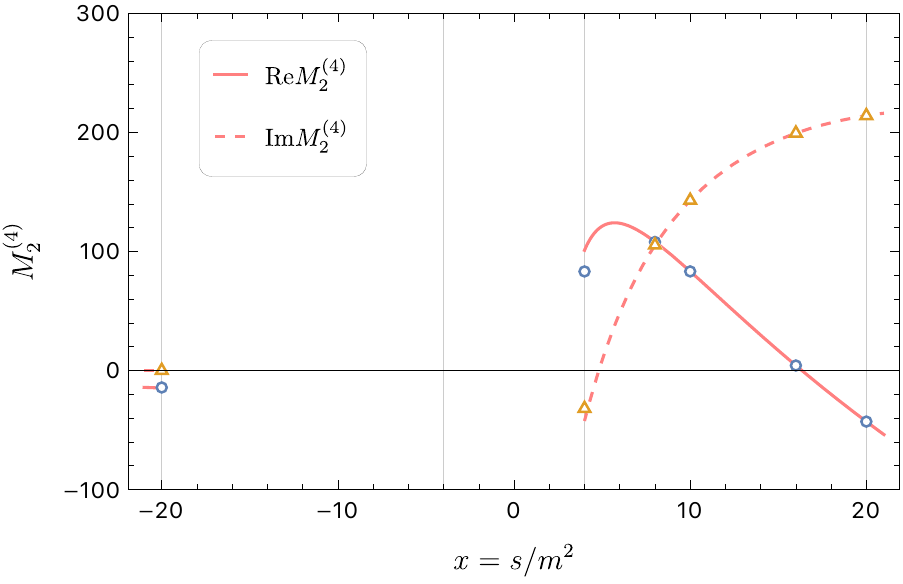}	
	\caption{\label{fig:appM1}Numeric results from $q$-expansion (solid and dashed lines) for the integrals $M_1$ and $M_2$ of family (a) in the region. Left: a broad range for $|x| > 20$; right: a closer look at $4 < |x| < 20$. The results are in good agreement with those from \texttt{AMFlow} (circles and triangles). }
\end{figure}

In Fig.~\ref{fig:appM1}, we show the numeric results from $q$-expansion for the integrals $M_1$ and $M_2$ in the region $x \in (-\infty, -20) \cup (4,\infty)$. We give the first two orders of $\varepsilon$, $M_1=\sum_{n\geq 4}\varepsilon^n M_1^{(n)}$ and $M_2=\sum_{n\geq 3}\varepsilon^n M_2^{(n)}$. The $q$-expansion converges rather fast in this region. With the first 8 orders in the expansion, our results perfectly agree with those computed using \texttt{AMFlow}~\cite{Liu:2017jxz,Liu:2020kpc,Liu:2022chg}. For example, for all the functions we plot here, there is a relative error around $\mathcal{O}(10^{-8})$ at $x=-50$ and a relative error around $\mathcal{O}(10^{-10})$ at $x=50$.

For regions of $y$ outside $(-1/4, 1/20)$, we note that in the case of MPLs, one may employ various transformations \cite{Vollinga:2004sn} to bring them back to the regions where series expansions converge. In the elliptic cases, we will need to employ the modular transformations in addition. We leave this to future investigations.

\section{The top sector integrals in family (b)}
\label{sect:familyb}

We now turn to the top sector of family (b). Our treatment is essentially similar to family (a), with some new ingredients. These ingredients are generic and appear in other cases as well. The first issue is that there are more than two MIs in the top sector, and we need to introduce new objects in the ansatz of canonical MIs. Secondly, the dependence on the sub-sector integrals is not automatically $\varepsilon$-factorized as the inhomogeneous term of the Picard-Fuchs equation. We introduce a subtraction scheme, which can be applied to other cases, to achieve factorization.

\subsection{The Picard-Fuchs operator and $\varepsilon$-factorization}
\label{sect:ansatzb}

\subsubsection{The maximal cut}
\label{sect:top}

There are three MIs in the top sector. The first-order differential equations for them are equivalent to the following third-order differential equation for $I_1$:
\begin{equation}
	\label{eq:PFepsb}
	\begin{aligned}
	L_3^{(\varepsilon)}(y) \, I_1 = \sum_{i=0}^3 r_i(y, \varepsilon)\frac{d^i}{d y^i}   I_1 = R(y,\varepsilon) \,,
	\end{aligned}
\end{equation}
where the inhomogeneous terms on the right-hand side are sub-sector integrals and will be discussed later. The coefficients in the Picard-Fuchs operator are
\begin{equation}
	\label{eq:PFepscoeffsb}
	\begin{aligned}
		r_3(y, \varepsilon) &= 1\,,\\
		r_2(y, \varepsilon) &= \frac{6 y \big[(1+4y) \varepsilon -1\big]+3}{y (y+1) (8 y-1)}\,,\\
		r_1(y, \varepsilon) &= \frac{\left(16 y^3+12 y^2+4\right) \varepsilon ^2+\left(-24 y^3-18 y^2-12 y\right) \varepsilon +8 y^3+30 y^2+6 y-7}{y^2 (y+1)^2 (8 y-1)}\,,\\
		r_0(y, \varepsilon) &= \frac{1}{y^3 (y+1)^2 (8 y-1)}\left[-(  8 y^2+8 y) \varepsilon^3 -(16 y^3+4 y^2+8 y+8) \varepsilon^2\right.\\
		&\left.\quad+(24 y^3+20 y^2+14 y)\varepsilon -8 y^3-32 y^2-4 y+8\right].
	\end{aligned}
\end{equation}
Although this operator is of order 3, when $\varepsilon=0$, it factorizes into the composition of a second-order operator $L_2^{(0)}(y)$ and a first-order operator $L_1^{(0)}(y)$, found via \texttt{DFactor}~\cite{van2012galois} in \texttt{Maple}:
\begin{equation}
	\label{eq:opfact}
	\begin{aligned}
		L_3^{(0)}(y) &= \underbrace{ \left[\frac{d}{d y}+\frac{8}{8y-1}\right]}_{L_1^{(0)}(y)} \underbrace{\left[\frac{d^2}{d y^2}+\left(\frac{1}{y+1}+\frac{8}{8 y-1}-\frac{3}{y}\right)\frac{d}{d y}+\frac{8 y (y+2)-4}{y^2 (y+1) (8 y-1)}\right]}_{L_2^{(0)}(y)} \, .
	\end{aligned}
\end{equation}
The irreducible second-order operator $L_2^{(0)}(y)$ is associated with an elliptic curve similar to that in family (a). This factorized structure appears very often in elliptic integral families and will simplify the situation significantly. We illustrate this simplification in the following.

In general, the complexity of the underlying geometric object increases with the order of the Picard-Fuchs operator when it is irreducible. For example, there are cases associated with hyperelliptic curves \cite{Georgoudis:2015hca} or Calabi-Yau $n$-folds \cite{Brown:2010bw, Bourjaily:2018yfy, Bourjaily:2019hmc, Klemm:2019dbm, Bonisch:2020qmm, Duhr:2022pch, Duhr:2022dxb}. However, when the operator is factorable, like in Eq.~\eqref{eq:opfact}, the geometry becomes simpler. In this case, although the operator is a third-order one, the essence is still an elliptic curve instead of more complicated objects. One may generalize the above observation to higher-order operators. Suppose that an $(n+1)$-order operator $L_{n+1}^{(0)}(y)$ factorizes into
\begin{equation}
	\label{eq:opfactgeneral}
	L_{n+1}^{(0)}(y) = \left[\frac{d}{d y}+\frac{1}{y+a}\right] L_{n}^{(0)}(y) \, .
\end{equation}
We can expect that the geometry is determined by the $n$-order operator $L_{n}^{(0)}(y)$. In fact, these two operators are equivalent, up to an inhomogeneous term. We call the solution space of a differential operator the space spanned by its linearly independent solutions. Then the solution space of $L_{n+1}^{(0)}(y)$, whose dimension is $n+1$, can be decomposed into a $n-$dimensional subspace, which is the solution space of $L_{n}^{(0)}(y)$ (spanned by the solutions $\{\psi_i\}$ ($i = 0,1,\ldots,n-1$) for $L_{n}^{(0)}(y)\psi_i=0$) and an extra solution $\psi_n$ determined by
\begin{equation}
	\label{eq:opfactgeneralcont}
	L_{n}^{(0)}(y)\psi_{n}(y) = \frac{c}{y+a} \,,
\end{equation}
where $c$ is a constant, i.e., the extra solution annihilated by $L_{n+1}^{(0)}(y)$ is essentially a special solution to $L_{n}^{(0)}(y)$.

The operators $L_3^{(0)}(y)$ and $L_2^{(0)}(y)$ have four regular singular points: $\{-1,\,0,\, 1/8,\,\infty\}$. Again, $0,\,\infty$ are the two points where the solutions have uniform asymptotic behaviors. Similar to family (a), we focus on the point $y=0$ here. The Frobenius method tells us to write the solutions of $L_3^{(0)}(y)$ as
\begin{equation}
	\label{eq:periodsdefb}
	\begin{aligned}
		\psi_{k}(y) = \frac{1}{(2 \pi i)^{k}} \sum_{j=0}^{k} \frac{\ln ^{j} y}{j !} \sum_{n=0}^{\infty} a_{k-j, n} \, y^{n+2} \,, \quad (k=0,1,2) \,.
	\end{aligned}
\end{equation}
The overall normalization is fixed by $a_{0,0}=1$. The first solution $\psi_0$ is holomorphic, while the other two have logarithmic behaviors. The expansion coefficients can be obtained iteratively, and the results read
\begin{equation}
	\label{eq:periodsb}
	\begin{aligned}
		\psi_0(y) &= y^2\left(1 + 2 y + 10 y^2 + 56 y^3 + 346 y^4 + 2252 y^5\right)+\mathcal{O}(y^8),\\
		\psi_1(y) &= \frac{1}{2\pi i} \left[\psi_0\ln y + \frac{y^3}{20}\left(60 + 330 y + 2000 y^2 + 12805 y^3 + 85262 y^4\right) + \mathcal{O}(y^8)\right],\\
		\psi_2(y) &= \frac{1}{(2\pi i)^2}\left[-\psi_0\frac{\ln^2y}{2}+2\pi i\,\psi_1\ln y+\frac{3 y^4}{16}\left(12 + 120 y + 931 y^2 + 6910 y^3\right)+\mathcal{O}(y^8)\right].
	\end{aligned}
\end{equation}
One can check that $\psi_{0}$ and $\psi_1$ are annihilated by $L_2^{(0)}$, i.e., they are associated with the two independent periods of the underlying elliptic curve. We again define the modular variable $\tau \equiv \psi_1/\psi_0$ and $q \equiv \exp(2\pi i\tau)$. The relation between $y$ and $q$ (around $y=0$) is given by
\begin{equation}
	\label{eq:qyb}
	\begin{aligned}
		y(q) &= q - 3 q^2 + 3 q^3 + 5 q^4 - 18 q^5 + 15 q^6 + 24 q^7+\mathcal{O}(q^8) \,,
		\\
		q(y) &= y + 3 y^2 + 15 y^3 + 85 y^4 + 522 y^5 + 3366 y^6 + 22450 y^7 + \mathcal{O}(y^8) \,.
	\end{aligned}
\end{equation}
The $q$-expansion of $y$ in Eq.~\eqref{eq:qya} can be expressed in terms of Dedekind eta-quotients, and we find that
\begin{equation}
	\label{eq:eta}
	y(\tau) = \frac{\eta(\tau)^3 \, \eta(6\tau)^9}{\eta(2\tau)^3 \, \eta(3\tau)^9} \,,
\end{equation}
where $\eta(\tau)$ is the Dedekind eta function. The above series can also be found in the Online Encyclopedia of Integer Sequences (OEIS) \cite{OEIS:A123633}. We will extend the second line by studying the associated elliptic curve later on, such that given a $y\in \mathbb{R}+i0$, we obtain a $\tau$ via $\tau(y)$ and then return the same $y$ via $y\big(\tau(y)\big)$ without ambiguity. The rest of this subsection does not rely on this but on the requirement that $\psi_i$ are periods of the Picard-Fuchs operator. As a result, the construction is as general as possible.

The Jacobian $J(y)$ for the variable change from $y$ to $\tau$ or $q$ is defined similarly to Eq.~\eqref{eq:Jacobian}, and can be expressed as
\begin{equation}
	\label{eq:Jacobianb}
	J(y) = \frac{\psi_0^2(y)}{W(y)} \, ,
\end{equation}
where the Wronskian is now
\begin{equation}
	\label{eq:Legendre}
	W(y) \equiv 2\pi i\left[\psi_{0} \frac{d}{d y} \psi_{1}-\psi_{1} \frac{d}{d y} \psi_{0}\right]=\frac{y^{3}}{(1-8 y)(1+y)} \,.
\end{equation}
It's easy to check that $\psi_2$ satisfies
\begin{equation}
	L_2^{(0)}(y) \, \psi_2(y) = \frac{1}{1-8y} \, ,
\end{equation}
which is an example of Eq.~\eqref{eq:opfactgeneralcont}. We can also rewrite $\psi_2$ as an iterated integral with $\psi_0$ and $\tau$ as the functional degrees of freedom:
\begin{equation}
	\label{eq:psi2}
	\begin{aligned}
		\psi_2(\tau) = \psi_0(\tau) \int\limits_{i\infty}^{\tau} d\tau_2 \int\limits_{i\infty}^{\tau_2} d\tau_1 \, \frac{(1-8y_1)(1+y_1)^2}{y_1^6}\, \psi_0(\tau_1)^3 \,,
	\end{aligned}
\end{equation}
where $y_1=y(\tau_1)$.

We now want to factorize the operator $L_3^{(0)}$ completely in $q$-space. Although $L_3^{(0)}$ is reducible and is not a genuine Calabi-Yau operator, we can borrow the treatment for the latter and define the so-called ``{\it normal forms}''~\cite{Bogner:13, Pogel:2022vat, Duhr:2022dxb}:
\begin{equation}
	\label{eq:normforms}
	\begin{aligned}
	   \alpha_{1}(y) & =\left[y \frac{d}{d y} \frac{\psi_{1}}{\psi_{0}}\right]^{-1}=\frac{J(y)}{y} \,, \\
	   \alpha_{2}(y) & =\left[y \frac{d}{d y} \alpha_{1}(y) y \frac{d}{d y} \frac{\psi_{2}}{\psi_{0}}\right]^{-1}=\left[\frac{y}{J(y)} \frac{d^{2}}{d \tau^{2}} \frac{\psi_{2}}{\psi_{0}}\right]^{-1} \,.
	\end{aligned}
\end{equation}
From the above we can define \textit{one} ``$Y$-invariant'':
\begin{equation}
	\label{eq:Yinv}
	\begin{aligned}
		Y(\tau) = \frac{\alpha_1}{\alpha_2} = \frac{d^2}{d\tau^2}\frac{\psi_2}{\psi_0}
		=\frac{\psi_0(y)^3}{(1-8y)W(y)^2} \,.
	\end{aligned}
\end{equation}
Interestingly, we find a compact expression for this quantity with the Lambert series:
\begin{equation}
	\label{eq:LambertY}
	Y(\tau) =1+\sum_{n=1}^{\infty} a(n) \frac{q^{n}}{1-q^{n}},\quad \text{with}\quad  a(n)=\left\{ \begin{aligned}
-\left(9 \left\lfloor \frac{n}{6}\right\rfloor+3\right)^{2},& \quad n\equiv 2\,\,({\rm mod}\,6),\\
\left(9 \left\lfloor \frac{n}{6}\right\rfloor+6\right)^{2},&\quad n\equiv 4\,\,({\rm mod}\,6),\\
0,&\quad {\rm otherwise},
\end{aligned} \right.
\end{equation}
where $\lfloor x\rfloor$ stands for the floor function~\footnote{We thank David Broadhurst for encouraging us to pursue this Lambert-series representation.}.  With the help of the $Y$-invariant, we can write the Picard-Fuchs operator in the factorized form in $q$-space:
\begin{equation}
	\label{eq:PFfacb}
	L_3^{(0)}(y) = \frac{\psi_0 Y}{J^3}\Theta_q \frac{1}{Y}\Theta_q^2 \frac{1}{\psi_0} \, .
\end{equation}
It is easy to check that the right-hand side annihilates $\psi_0$, $\psi_1$, and $\psi_2$ as expected.

The above factorization pattern is essential to write down the ansatz for the canonical MIs in the top sector, and the workflow can be streamlined for other elliptic integral families. Following \cite{Pogel:2022ken, Pogel:2022yat}, we write the ansatz as:
\begin{equation}
	\label{eq:ansatzb}
	\begin{aligned}
		M_1 &= \varepsilon^4\frac{I_{1}}{\psi_0}\,,\\
		M_2 &= \frac{J(y)}{\varepsilon}\frac{d}{d y} M_1 - F_{11} M_1\,,\\
		M_3 &= \frac{1}{Y(y)}\left[\frac{J(y)}{\varepsilon}\frac{d}{d y} M_2 - F_{21} M_1 - F_{22}M_2\right] + \text{sub-sector integrals} \,.
	\end{aligned}
\end{equation}
We demand that the differential equations for the above basis are $\varepsilon$-factorized under maximal cut (so the sub-sectors integrals are irrelevant at the moment). This leads to constraints on the coefficient functions $F_{ij}$.

Under maximal cut, the differential equation for the above ansatz takes the following form:
\begin{equation}
	\label{eq:Ansatzde}
	\begin{aligned}
		J(y)\frac{d}{d y}
		\begin{pmatrix}
			M_1\\
			M_2\\
			M_3
		\end{pmatrix}^{\text{mc}} = A^{\text{mc}}
		\begin{pmatrix}
			M_1\\
			M_2\\
			M_3
		\end{pmatrix}^{\text{mc}}
		=
		\begin{pmatrix}
			\varepsilon\,F_{11} & \varepsilon & 0\\
			\varepsilon\,F_{21} & \varepsilon\,F_{22} & \varepsilon\,Y\\
			A_{31} & A_{32} & A_{33}
		\end{pmatrix}
		\begin{pmatrix}
			M_1\\
			M_2\\
			M_3
		\end{pmatrix}^{\text{mc}} .
	\end{aligned}
\end{equation}
The first and second rows of $A^{\text{mc}}$ are automatically $\varepsilon$-factorized. We now need to bring the third row to the $\varepsilon$-form by appropriate choices of the coefficient functions. We first note that $J(y)$ and $Y(y)$ as in Eqs.~\eqref{eq:Jacobianb} and \eqref{eq:Yinv} satisfy the following relation:
\begin{equation}
	\label{eq:constraint1}
	\begin{aligned}
		3\frac{J'(y)}{J(y)} = 3\frac{\psi'_0(y)}{\psi_0(y)}+\frac{Y'(y)}{Y(y)} + \frac{1}{y+1}+\frac{16}{8 y-1}-\frac{3}{y} \,,
	\end{aligned}
\end{equation}
which makes the $\mathcal{O}(\varepsilon^0)$ terms in $A_{33}$ vanish.
Combining $L_2^{(0)}\psi_0=0$ and Eq.~\eqref{eq:Yinv}, we can see that $A_{32}$ does not contain $\varepsilon^{-1}$ terms. The absence of $\varepsilon^{-1}$ terms in $A_{31}$ can be used to constrain the coefficient function $F_{11}$. It is convenient to rewrite $F_{11}(y)$ as \footnote{This is inspired by \cite{Pogel:2022yat} and is in agreement with the algorithm in \cite{broedel2018modular}.}
\begin{equation}
	\label{eq:F11refine}
	F_{11}(y) = F(y) +\frac{28 y^2+2 y+1}{3 y^4} \, \psi_0(y)^2 \,.
\end{equation}
To remove the $\varepsilon^{-1}$ terms in $A_{31}$, the function $F(y)$ must satisfy the constraint
\begin{equation}
	\label{eq:constraint3}
	\begin{aligned}
		F''(y)+\left[\frac{J'(y)}{J(y)}-\frac{Y'(y)}{Y(y)}\right]F'(y) = 0 \, .
	\end{aligned}
\end{equation}
The solution reads
\begin{equation}
	\label{eq:solF}
	F(y) = c_1 + c_2\int\limits_{i\infty}^{\tau} d\tau_1 \, Y(\tau_1) = c_1 + \frac{c_2}{2\pi i} \int\limits_{0}^{q} \frac{dq_1}{q_1} \, Y(q_1) \, ,
\end{equation}
where $c_1$ and $c_2$ are integration constants. Since $Y(q) = 1 + \mathcal{O}(q^2)$, the second term in the above expression is logarithmic divergent. To avoid that, we choose $c_2=0$. The constant $c_1$ is arbitrary and for convenience we also set it to zero. Consequently, we obtain
\begin{equation}
	\label{eq:F11}
	F_{11}(y) = \frac{28 y^2+2 y+1}{3 y^4} \, \psi_0(y)^2 \,.
\end{equation}
The function $F_{22}(y)$ can be used to remove the $\varepsilon^{0}$ terms in $A_{32}$, with the following constraint:
\begin{equation}
	\label{eq:constraint4a}
	\begin{aligned}
		F'_{22}(y)-\frac{Y'(y)}{Y(y)} F_{22}(y)  = \left[\frac{112 y^4+308 y^3-45 y^2+y-1}{y^5 (y+1) (8 y-1)}-\frac{28 y^2+2 y+1}{3 y^4}\frac{\psi'_0(y)}{\psi_0(y)}\right]\psi_0(y)^2 \,.
	\end{aligned}
\end{equation}
It is easy to verify that $F_{22}(y)=F_{11}(y)$ in \eqref{eq:F11} is a solution to the above equation. Further demanding $A_{31}$ to be free from $\varepsilon^{0}$ terms, we have the following constraint on $F_{21}(y)$:
\begin{equation}
	\label{eq:constraint4b}
	\begin{aligned}
		F'_{21}(y)-\frac{Y'(y)}{Y(y)} F_{21}(y) &= \left[\frac{(2 y-1) \left(88 y^3+84 y^2+66 y-11\right)}{3 y^8}\frac{\psi'_0(y)}{\psi_0(y)}\right.\\
		&\left.-\frac{2 \left(704 y^6-64 y^5+712 y^4-1224 y^3-352 y^2+154 y-11\right)}{3 y^9 (y+1) (8 y-1)}\right]\psi_0(y)^4 \,.
	\end{aligned}
\end{equation}
A specific solution reads
\begin{equation}
	\label{eq:F21sol}
	F_{21}(y) = \frac{(2 y-1) \left(88 y^3+84 y^2+66 y-11\right)}{3 y^8} \, \psi_0(y)^4 \,.
\end{equation}
Collecting all the entries, the $3 \times 3$ connection matrix in the top sector reads
\begin{equation}
	\label{eq:epsformtopb}
	\begin{aligned}
	    \setlength{\arraycolsep}{5pt}
	    \renewcommand*{\arraystretch}{2}
		A^{\text{mc}} = \varepsilon\,
		\begin{pmatrix}
		\frac{\left(28 y^2+2 y+1\right) \psi_0 (y)^2}{3 y^4} & 1 & 0 \\
		\frac{(2 y-1) \left(88 y^3+84 y^2+66 y-11\right) \psi_0 (y)^4}{3 y^8} & \frac{\left(28 y^2+2 y+1\right) \psi_0 (y)^2}{3 y^4} & -\frac{(y+1)^2 (8 y-1) \psi_0 (y)^3}{y^6} \\
		\frac{64 (y+1)^2 (8 y-1) \psi_0 (y)^3}{27 y^6} & 0 & \frac{2 (y+1) (8 y-1) \psi_0 (y)^2}{3 y^4}
		\end{pmatrix} .
	\end{aligned}
\end{equation}

The structure of the above matrix hints us to perform a further rotation of the master integrals:
\begin{equation}
	\label{eq:refine}
	\begin{aligned}
		\begin{pmatrix}
			M'_1\\
			M'_2\\
			M'_3
		\end{pmatrix}=\begin{pmatrix}
			1 & 0 & 0\\
			0 & 0 & \frac{3\sqrt{3}\,i}{8}\\
			0 & 1 & 0
		\end{pmatrix}
		\begin{pmatrix}
			M_1\\
			M_2\\
			M_3
		\end{pmatrix},
	\end{aligned}
\end{equation}
The differential equation for the new basis takes a similar form as Eq.~\eqref{eq:Ansatzde}:
\begin{equation}
	\label{eq:epsformtopbnew}
	\begin{aligned}
		J\frac{d}{d y}\begin{pmatrix}
			M'_1\\
			M'_2\\
			M'_3
		\end{pmatrix}^{\text{mc}} = \varepsilon\,
		\begin{pmatrix}
		\eta'_{1,2} & 0 & 1 \\
		\eta'_3 & \eta'_{2,2} & 0 \\
		\eta'_4 & \eta'_{3} & \eta'_{1,2}
		\end{pmatrix}\begin{pmatrix}
			M'_1\\
			M'_2\\
			M'_3
		\end{pmatrix}^{\text{mc}} ,
	\end{aligned}
\end{equation}
The new connection matrix is persymmetric and of uniform modular weight, with the entries
\begin{equation}
	\label{eq:letterstop}
	\begin{aligned}
		\eta'_{1,2} &= \frac{28y^2+2y+1}{3y^4}\psi_0(y)^2 \,,\\
		\eta'_{2,2} &= \frac{2(y+1)(8y-1)}{3y^4}\psi_0(y)^2 \,,\\
		\eta'_{3}   &= \frac{8 i (y+1)^2 (8 y-1) \psi_0 (y)^3}{3 \sqrt{3} y^6}=\frac{8i}{3\sqrt{3}}Y \,,\\
		\eta'_4     &= \frac{(2 y-1) \left(88 y^3+84 y^2+66 y-11\right) \psi_0 (y)^4}{3 y^8} \,.
	\end{aligned}
\end{equation}
However, this new basis is inconvenient in practice, and we won't use it in the following.

\subsubsection{Including sub-sector independence}
\label{sect:sub}

The inhomogeneous term in \eqref{eq:PFepsb} reads:
\begin{equation}
	\label{eq:PFepsbinhomo}
	\begin{aligned}
		R(y,\varepsilon) = \left[\varepsilon^2 \vec{f}_{\rm 2, sub}(y) + \varepsilon^3 \vec{f}_{\rm 3, sub}(y)  \right]\cdot \vec{M}_{\rm sub},
	\end{aligned}
\end{equation}
with $\vec{M}_{\rm sub}=(M_4,\,M_5,\cdots,M_{18})^{T}$ and
\begin{equation}
	\label{eq:PFepsbinhomocont}
	\begin{aligned}
		\vec{f}_{\rm 2, sub}(y) =& \frac{-4}{(1+y)^2(8y-1)}\bigg(0,-5,5,4,4,\frac{16 y^2-58 y+1}{-2 (1-4 y)^{3/2}},\frac{9}{2},-5,-6,-4,2,\frac{7}{2},29,15,4\bigg),\\
		\vec{f}_{\rm 3, sub}(y) =& \frac{4}{y(1+y)^2(8y-1)}\bigg(0, -3 + 7 y, 3 - 7 y, -8 y, -8 y, -\frac{(y+1) \left(48 y^2-4 y+3\right)}{(1-4 y)^{3/2}},\\
		&\frac{3 (y+1) (2 y-3)}{4 y-1}, \frac{13 y^2+10 y+5}{y-1},-2 (y+1),-2 (8 y+3),\frac{5 \left(8 y^2-y+1\right)}{2 (4 y-1)},\\
		&\frac{1}{2} (-3 y-1),\frac{35 y^2+66 y+11}{y-1},\frac{(y+1) (13 y+15)}{y-1},-\frac{128 y^3+115 y^2+18 y-9}{2 (y-1) (4 y-1)} \bigg).
	\end{aligned}
\end{equation}

Apparently, $R(y,\varepsilon)$ is not $\varepsilon$-factorized due to the presence of $\vec{f}_{\rm 2, sub}$. This leads to non-factorized $\varepsilon$-dependence in the differential equations for $M_3$, because this inhomogeneous term enters the differential equation by substituting the third-order derivative by lower ones. It turns out that it is $\vec{f}_{\rm 2, sub}$ that makes the $M_3$'s sub-sector dependence not $\varepsilon$-factorized. A systematic and minimal prescription to eliminate this is subtracting certain sub-sector contributions in the definition of $M_3$, as hinted in Eq.~\eqref{eq:ansatzb}. Here we write, \textit{a priori},
\begin{equation}
	\label{eq:M3newdef}
	M_3 = \frac{1}{Y(y)}\left[\frac{J(y)}{\varepsilon}\frac{d}{d y} M_2 - F_{21}(y) M_1 - F_{22}(y)M_2\right]-\vec{g}_{\rm 2, sub}(y)\cdot \vec{M}_{\rm sub} \,.
\end{equation}
Note that $\vec{M}_{\rm sub}$ is already a UT basis for the sub-sectors. Substitute the refined $M_3$ into the differential equation, and one finds that the following constraint:
\begin{equation}
	\label{eq:subtractioncondition}
	\begin{aligned}
		\frac{d}{d y} \vec{g}_{\rm 2, sub}(y) = (1-8y) \vec{f}_{\rm 2, sub}(y) \,,
	\end{aligned}
\end{equation}
can cancel the contribution from $\vec{f}_{\rm 2, sub}$ in the inhomogeneous terms. It is easy to find a particular solution to the above equation, and we choose it to be
\begin{equation}
	\label{eq:subtraction}
	\begin{aligned}
		\vec{g}_{\rm 2, sub}(y) = \frac{2}{1+y}\left(0,10,-10,-8,-8,\frac{7-8 y}{\sqrt{1-4 y}},-9,10,12,8,-4,-7,-58,-30,-8\right).
	\end{aligned}
\end{equation}
Now, the differential equations for the three top-sector integrals are given by
\begin{equation}
    \label{eq:topepsb}
	\begin{aligned}
	J(y) \frac{\mathrm{d}}{\mathrm{d}y} M_{1} &= \varepsilon\left[F_{11}(y)M_{1}+M_{2}\right] , \\
	J(y) \frac{\mathrm{d}}{\mathrm{d}y} M_{2} &= \varepsilon\left[F_{21}(y)M_{1}+F_{22}(y)M_{2}+Y(y)M_{3}+Y(y) \, \vec{g}_{\rm 2, sub}(y)\cdot \vec{M}_{\rm sub} \right] ,\\
	J(y) \frac{\mathrm{d}}{\mathrm{d}y} M_{3} &= \varepsilon \, \Big[ A_{31}(y)M_{1}+A_{32}(y)M_{2}+A_{33}(y)M_{3}+J(y) \, \vec{g}_{\rm 3, sub}(y)\cdot \vec{M}_{\rm sub} \Big] \, ,
	\end{aligned}
\end{equation}
where $A_{31}(y)$, $A_{32}(y)$ and $A_{33}(y)$ have been derived in \eqref{eq:epsformtopb} and
\begin{equation}
	\label{eq:f3tilde}
	\begin{aligned}
		\vec{g}_{\rm 3, sub}(y) = \frac{16}{3y(y+1)}\left(0,-4,4,2,2,-\frac{4 (y+1)}{\sqrt{1-4 y}},0,\frac{4 y+2}{1-y},0,4,-2,1,-\frac{8 (y+2)}{y-1},\right.\\
		\left.-\frac{6 (y+1)}{y-1},\frac{5 y+4}{y-1}\right),
	\end{aligned}
\end{equation}
is the consequence of the subtraction term and the remaining inhomogeneous term.

\subsection{Elliptic curve and analytic continuation of the modular map}

\begin{figure}[t!]
  \centering
  \includegraphics[width=0.4\textwidth]{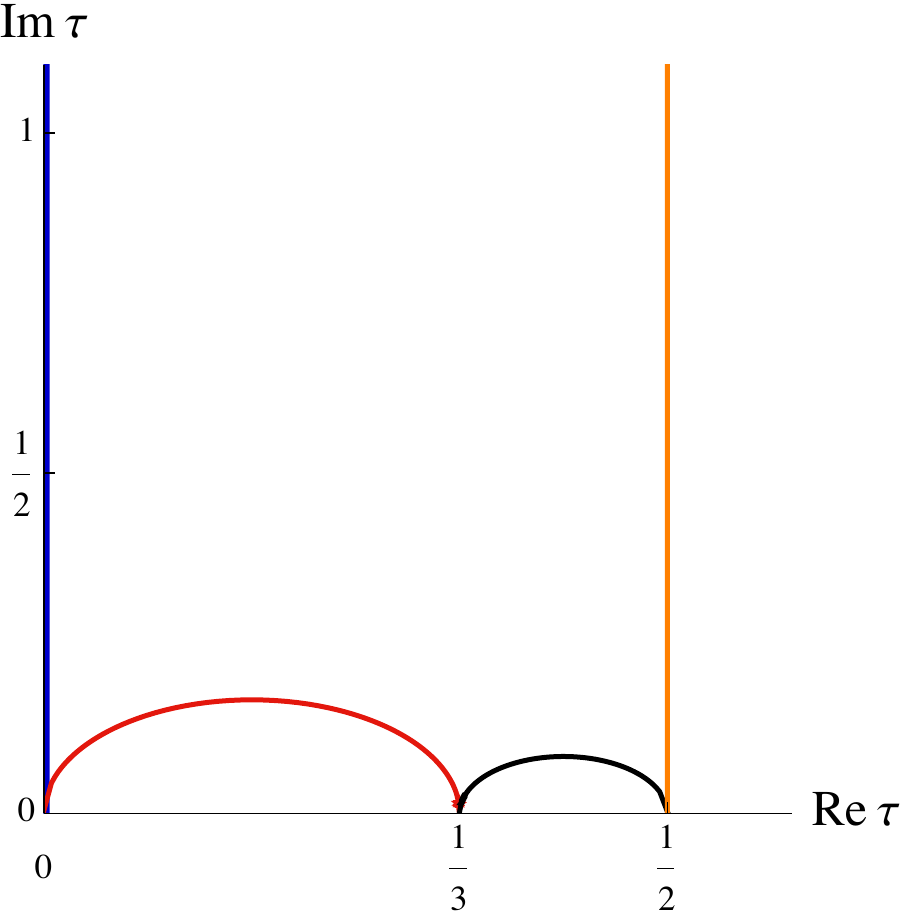}
  \hspace{1cm}
  \includegraphics[width=0.4\textwidth]{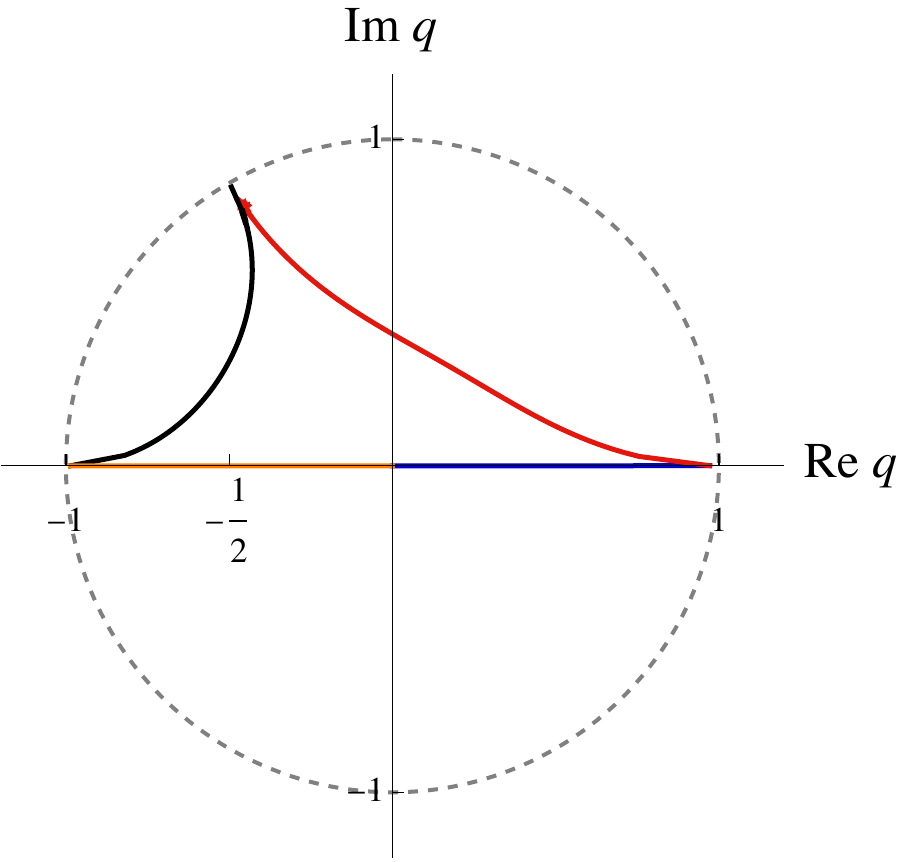}
  \caption{\label{fig:pathb}The behaviors of $\tau$ and $q$ as functions of $y$. The black path corresponds to $y\in (-\infty, -1)$, the orange one corresponds to $y\in [-1,0)$, the blue one corresponds to $y\in [0,\,1/8)$, and the red one corresponds to $y\in [1/8,+\infty) $. }
\end{figure}

\begin{figure}[t!]
  \centering
  \includegraphics[width=10cm]{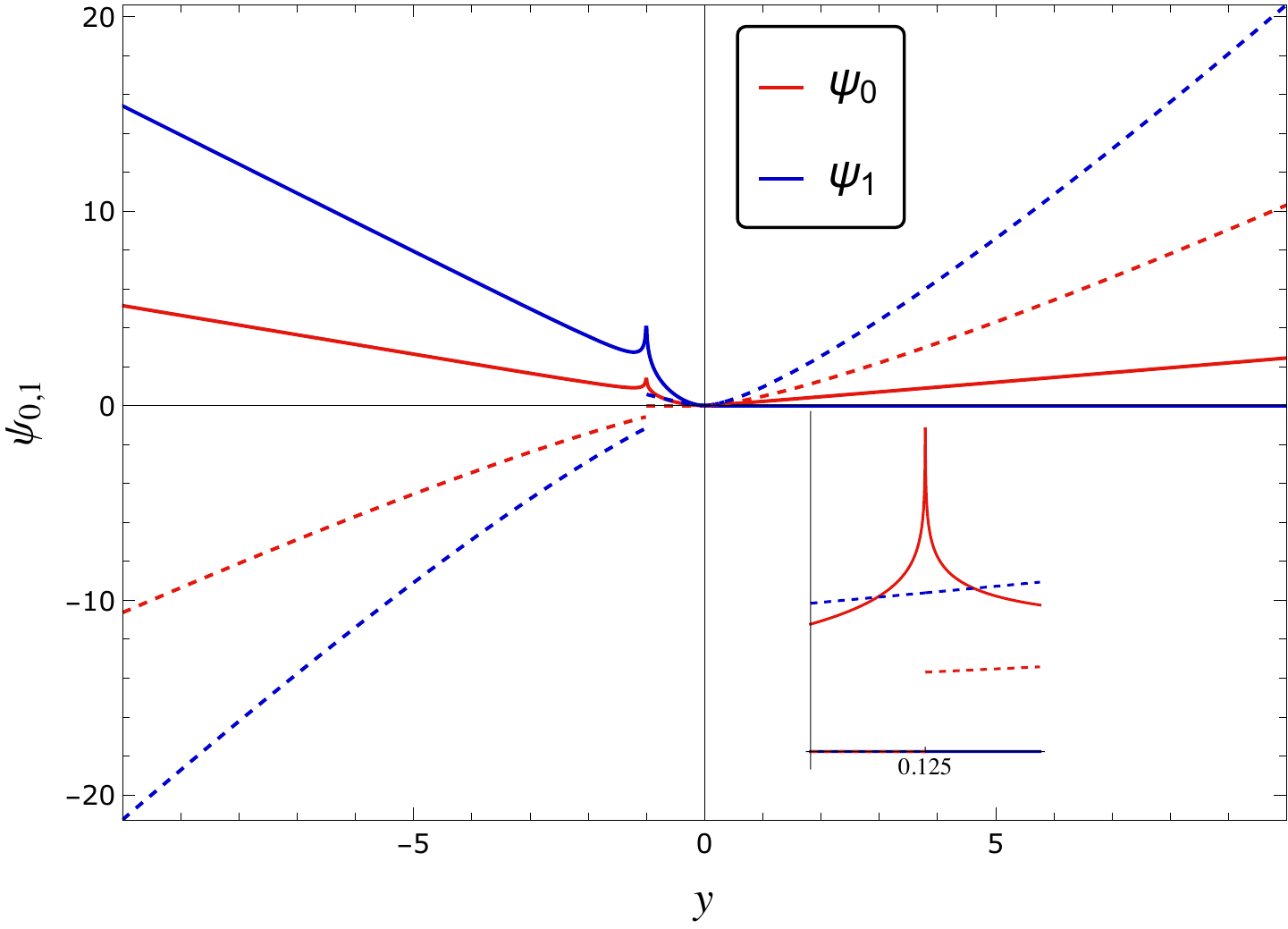}
  \caption{\label{fig:periodsb}The two periods after the analytic continuation. The solid lines are real parts, and the dashed lines are imaginary parts. There are four cusps for $\Gamma_1(6)$: $\tau=0,\,1/3,\,1/2,\,i\infty$. The first period $\psi_0$ is holomorphic around $\tau=i\infty$ ($y=0$), but is not continuous around the cusp $\tau=1/2$ ($y=-1$), and the cusp $\tau=0$ ($y=1/8$), which is shown as a sub-plot in the above.}
\end{figure}

As in the case of family (a), we'd like to extend the relation between $y$ and $\tau$ to the entire phase space, given by Eq.~\eqref{eq:eta}. For that purpose, we study the associated elliptic curve, obtained by the maximal cut of $I_1$:
\begin{equation}
	\label{eq:Eb}
		E: \; v^2 = \left(u-u _1\right)\left(u-u_2\right)\left(u-u_3\right)\left(u-u_4\right),
\end{equation}
with the four roots
\begin{equation}
	u_1 = -1 \,, \quad u_2 = -\frac{\left(\sqrt{1-8y}+1\right)^2}{4} \,, \quad u_3 = -\frac{\left(\sqrt{1-8y}-1\right)^2}{4} \,, \quad u_4 = 0 \,.
\end{equation}
In the region $0 < y < 1/8$, the four roots are ordered as $u_1<u_2<u_3<u_4$. We define the elliptic modulus as
\begin{equation}
	\label{eq:modulus}
	k^2 = \frac{(u_2-u_1)(u_4-u_3)}{(u_3-u_1)(u_4-u_2)} = \frac{1-4y-8y^2-\sqrt{1-8y}}{1-4y-8y^2+\sqrt{1-8y}} \,.
\end{equation}
Similar to the family $(a)$, we define two periods as linear combinations of two complete elliptic integrals:
\begin{equation}
	\label{eq:anacperiodsb}
	\begin{aligned}
		\begin{pmatrix}
			\bar{\psi}_1(y)\\
			\bar{\psi}_0(y)
		\end{pmatrix}=
		\frac{2}{\pi}\frac{y^2}{\sqrt{(u_3-u_1)(u_4-u_2)}} \, \gamma(y) \begin{pmatrix}
			i\,K(1-k^2)\\
			K(k^2)
		\end{pmatrix},
	\end{aligned}
\end{equation}
where the monodromy matrix is given by
\begin{equation}
	\label{eq:monodromy}
		\gamma(y) =
		\begin{cases}
		\begin{aligned}
		&\begin{pmatrix}
			1 & 2\\
			0 & 1
		\end{pmatrix}, \quad y < 0 \text{ or } y \geq  \frac{\sqrt{3}-1}{4} \,,
		\\
		&\begin{pmatrix}
			1 & 0\\
			0 & 1
		\end{pmatrix},\quad 0 \leq y < \frac{\sqrt{3}-1}{4} \,.
		\end{aligned}
		\end{cases}
\end{equation}
One can check that the rotation above compensates for the discontinuity of the complete elliptic integral when its argument crosses the branch cut. The crossing happens twice: one at $k^2=\infty$, and the other at $k^2=2$ which corresponds to $y=(\sqrt{3}-1)/4$. Performing the series expansion around $y = 0$ and comparing with Eq.~\eqref{eq:periodsb}, we can identify $\psi_0 = \bar{\psi}_0$ and $\psi_1 = \bar{\psi}_1/6$, and
\begin{equation}
	\label{eq:newtaub}
	\tau = \frac{\bar{\psi}_1(y)}{6\bar{\psi}_0(y)} \,.
\end{equation}
It is easy to verify that the above function is the inverse map of Eq.~\eqref{eq:eta}. It is also straightforward to check that the kinematic singular points are mapped to the cusps of the congruence subgroup $\Gamma_1(6)$:
\begin{equation}
   \label{eq:cuspsb}	
   \tau(y=-1) = \frac{1}{2},\quad \tau(y=0) = i\infty,\quad \tau(y=1/8) = 0,\quad \tau(y=\infty) = \frac{1}{3}.
\end{equation}
Behaviors of $\tau$ and $q$ as functions of $y$ in the upper half complex plane are shown in Fig.~\ref{fig:pathb}. The analytically continued periods are shown in Fig.~\ref{fig:periodsb}, which are smooth for all values of $y$ except at cusps.

\subsection{Letters and modular forms}
\label{sect:modularformsb}

The relevant modular group for family (b) is $\Gamma_1(6)$. Like family (a), the kinematic variable $y(\tau)$ is a modular function, and so are rational functions of $y$. For example, the Wronskian
\begin{equation}
    \label{eq:Wronskiantau}
	W(y) = \frac{y^{3}}{(1-8 y)(1+y)} = \frac{\eta(\tau)^{2} \, \eta(6 \tau)^{22}}{\eta(2 \tau)^{10} \, \eta(3 \tau)^{14}} \,,
\end{equation}
corresponds to the sequence A002508 \cite{OEIS:A002508} on OEIS.
Modular forms of $\Gamma_1(6)$ have been extensively studied in the literature about Feynman integrals. The building block $\psi_0$ lives in the 2-dimensional vector space of weight-1 modular forms. In terms of the Eisenstein basis $\{e_{1,1},\,e_{2,1}\}$, it can be written as
\begin{equation}
    \label{eq:modularbasisb}
	\psi_0 = y^2(2e_{1,1}+4e_{2,1}) \, .
\end{equation}
For our purpose, it is convenient to introduce another basis:
\begin{equation}
	\label{eq:modularbasisbnew}
	\begin{aligned}
		b_{1,1} =  \frac{1}{\sqrt{3}}\frac{\psi_0}{y}\,,\quad b_{2,1} =  \frac{1}{\sqrt{3}}\frac{\psi_0}{y^2} \,.
	\end{aligned}
\end{equation}

The dependence among the top-sector integrals themselves in the differential equations \eqref{eq:topepsb} involves the following modular form letters (the matrix elements in Eq.~\eqref{eq:epsformtopb}, and the modular weight is manifest by the power of $\psi_0$):
\begin{align}
		\eta_{1,2} &= 28b_{1,1}^2+2b_{1,1}b_{2,1}+b_{2,1}^2 \,, \nonumber \\
		\eta_{2,2} &= 16b_{1,1}^2 + 14b_{1,1}b_{2,1}-2b_{2,1}^2 \,, \nonumber \\
		\eta_{1,3} &= -3\sqrt{3}\big[8b_{1,1}^3+15b_{1,1}^2b_{2,1}+6b_{1,1}b_{2,1}^2-b_{2,1}^3\big] \,, \nonumber \\
		\eta_4 &= 3\big[176b_{1,1}^4+80b_{1,1}^3b_{2,1}+48b_{1,1}^2b_{2,1}^2-88b_{1,1}b_{2,1}^3+11b_{2,1}^4\big] \,.
		\label{eq:letterstopdecomp}
\end{align}
The dependence of the top-sector integrals on the sub-sector ones is complicated in the presence of a square root of the modular function $1-4y(\tau)$ and a simple pole at $y=1$. The Jacobian or the $ Y$-invariant cancels the spurious poles at $y=0$ and $y=-1$. Two new modular forms appearing in the mixing letters are given by:
\begin{equation}
	\label{eq:letterssubdecomp}
	\begin{aligned}
		\eta_{3,2} = \frac{16J(y)}{3y(y+1)} &= 16(b_{2,1}^2-8b_{1,1}b_{2,1}),\,\\
		\eta_{2,3} = \frac{2Y(y)}{y+1} &= 6\sqrt{3}(b_{2,1}^3-b_{2,1}^2b_{1,1}-8b_{2,1}b_{1,1}^2).
	\end{aligned}
\end{equation}
The dimension of weight-2 modular forms is 3, and the vector space is spanned by $\{\eta_{1,2},\,\eta_{2,2},\,\eta_{3,2}\}$, for example:
\begin{equation}
	\begin{aligned}
		\frac{J(y)}{y} = -\frac{3}{2}\eta_{2,2} \,, \quad \frac{J(y)}{1+y} = -\frac{3}{2}\eta_{2,2}-\frac{3}{16}\eta_{3,2} \,.
	\end{aligned}
\end{equation}
Three extra letters are not modular forms:
\begin{equation}
	\label{eq:extraletter}
	\begin{aligned}
		\varrho = \frac{7-8y}{\sqrt{1-4y}}\eta_{2,3}\,,\quad \vartheta = \frac{1+y}{\sqrt{1-4y}}\eta_{3,2}\,,\quad  \varpi = \frac{\eta_{3,2}}{y-1}\,.
		\end{aligned}
\end{equation}
They are well-behaved in a neighborhood of $y=0$ but not globally for all $y\in\mathbb{R}+i0$, either due to the branch cut of the square root or the pole at $y=1$, which does not map to a cusp. We rewrite the $\varepsilon$-factorized differential equations for the top-sector integrals as follows:
\begin{align}
	\label{eq:epsformbfinal}
		\frac{1}{2\pi i}\frac{d\,M_1}{d \tau}  =  \varepsilon \big[&\eta_{1,2}M_1+M_2\big],\notag\\
		\frac{1}{2\pi i}\frac{d\,M_2}{d \tau}  =  \varepsilon \Big[&\eta_{4}M_1+\eta_{1,2}M_2+\eta_{1,3}M_3+10\eta_{2,3}M_5-10\eta_{2,3}M_6-8\eta_{2,3}M_7-8\eta_{2,3}M_8\notag\\
		& +{\color{red}\varrho} M_9-9\eta_{2,3}M_{10}+10\eta_{2,3}M_{11}+12\eta_{2,3}M_{12}+8\eta_{2,3}M_{13}-4\eta_{2,3}M_{14}\notag \\
		& -7\eta_{2,3}M_{15}-58\eta_{2,3}M_{16}-30\eta_{2,3}M_{17}-8\eta_{2,3}M_{18}\Big],\\
		\frac{1}{2\pi i}\frac{d\,M_3}{d \tau}  =  \varepsilon \bigg[&-\frac{64}{27}\eta_{1,3}M_1+\eta_{2,2}M_3  -4\eta_{3,2}M_{5} + 4\eta_{3,2}M_6 + 2\eta_{3,2}M_7 + 2\eta_{3,2}M_8 \notag\\
		&-4{\color{red}\vartheta} M_9 -(4\eta_{3,2}+6{\color{red}\varpi})M_{11} + 4 \eta_{3,2}M_{13} -2\eta_{3,2}M_{14} +\eta_{3,2}M_{15} \notag \\
		&-(8\eta_{3,2}+24{\color{red}\varpi})M_{16} -(6\eta_{3,2}+12{\color{red}\varpi}) M_{17} + (5\eta_{3,2}+9{\color{red}\varpi}) M_{18}  \bigg]. \notag
\end{align}

Similar as family (a), letters with modular weight 2 are still algebraic functions of $y$, and can be cast into the $d\log$ representations:
\begin{equation}
	\label{eq:dloglettersb}
	\begin{aligned}
		\eta_{1,2}\cdot 2\pi i d\tau &= \frac{1}{3}d\log y-d\log (y+1)-\frac{1}{2} d\log (1-8 y)\,,\\
		\eta_{2,2}\cdot 2\pi i d\tau &= -\frac{2}{3}d\log y,\,\quad \eta_{2,2}\cdot 2\pi i d\tau = \frac{16}{3}\big(d\log y-d\log(1+y)\big)\,,\\
		\vartheta\cdot 2\pi i d\tau &= \frac{16}{3}d\log\frac{1-\sqrt{1-4y}}{1+\sqrt{1-4y}}=\frac{16}{3}d\log t\,,\\
		\varpi\cdot 2\pi i d\tau &= \frac{8}{3}\big[d\log(1-y)+d\log(1+y)-2\,d\log y\big]\,.
	\end{aligned}
\end{equation}

\subsection{Boundary conditions}
\label{scet:boundaryb}

The boundary point is again taken to be $y=0$, and we need to calculate the asymptotic behaviors of the canonical MIs in this limit. Results for the sub-sector integrals are collected in the Appendix. For the top sector, we again employ the Mellin-Barnes representation. The 4-fold integral representation of $I_1$ reads
\begin{equation}
	\label{eq:MBb}
	\begin{aligned}
		I_{1} &= e^{2\varepsilon\,\gamma_E} \int\prod_{i=1}^4\frac{d\,z_i}{2\pi i}\,y^{2+2\varepsilon+z_1}\Gamma(-z_3)\Gamma(-z_4)\Gamma(\varepsilon-z_2)\Gamma(-\varepsilon-z_4)\Gamma(1-\varepsilon+z_2)\\
		&\times\Gamma(-z_1+z_4)\Gamma(-\varepsilon-z_1+z_4)\Gamma(-2\varepsilon-z_1+z_3)\Gamma(-3\varepsilon-z_1+z_2)\\
		&\times\frac{\Gamma(1+2\varepsilon+z_1-z_3)\Gamma(-z_2+z_3)\Gamma(\varepsilon-z_2+z_3)\Gamma(\varepsilon-z_2+z_4)\Gamma(-2\varepsilon+z_3-z_4)}{\Gamma(-2\varepsilon-z_1)\Gamma(-3\varepsilon-z_1)\Gamma(-\varepsilon-z_2+z_3-z_4)\Gamma(-\varepsilon-z_1-z_2+z_3-z_4)} \,.
	\end{aligned}
\end{equation}
We perform the asymptotic expansion for $y \to 0$ and the series expansion in $\varepsilon$. Up to order $\varepsilon^2$, the result for $I_1$ is given by
\begin{equation}
	\label{eq:boundaryI1b}
	\begin{aligned}
		\frac{I_1}{y^2}\bigg|_{y\to 0} =& \, \frac{7 }{12}L_y^4-\zeta_2 L_y^2  -20 \zeta_3 L_y-\frac{31 \zeta_4}{2}+\varepsilon\bigg[\frac{3}{20}L_y^5-9\zeta_2L_y^3- 34\zeta_3L_y^2-\frac{125\zeta_4}{2}L_y\\
		& +40\zeta_2\zeta_3+8\zeta_5\bigg]+\epsilon^2\left[\frac{37}{360}L_y^6-2\zeta_2L_y^4-\frac{88\zeta_3}{3}L_y^3+\frac{13\zeta_4}{4}L_y^2\right. \\
		&\left. +\left(100\zeta_2\zeta_3-112\zeta_5\right)L_y+\frac{287}{4}\zeta_6+188\zeta_3^2\right]+\mathcal{O}(\varepsilon^3) \, ,
	\end{aligned}
\end{equation}
with $L_y=\log y$.

Using the definition in Eqs.~\eqref{eq:ansatzb} and \eqref{eq:M3newdef}, it is straightforward to derive the boundary conditions for the canonical MIs in the top sector:
\begin{align}
	\label{eq:familybBD}
		M_1|_{y\to 0}^{\rm boundary} &= \varepsilon^4 \left[\frac{7 }{12}L_q^4-\zeta _2 L_q^2-20 \zeta _3 L_q-\frac{31 \zeta _4}{2}\right]+\varepsilon^5 \bigg[\frac{3 }{20}L_q^5-9 \zeta _2 L_q^3-34 \zeta _3 L_q^2 \nonumber
		\\
		&\hspace{5cm}-\frac{125 \zeta _4 }{2}L_q+40 \zeta _2 \zeta _3+8 \zeta _5\bigg]+\mathcal{O}(\varepsilon^6)\,, \nonumber
		\\
		M_2|_{y\to 0}^{\rm boundary} &= \varepsilon^3\left[\frac{7 L_q^3}{3}-2 \zeta _2 L_q-20 \zeta _3\right]+\varepsilon^4\bigg[\frac{5 }{9}L_q^4-\frac{80}{3} \zeta _2 L_q^2-\frac{184 \zeta _3 L_q}{3}-\frac{172 \zeta _4}{3}\bigg]+\mathcal{O}(\varepsilon^5)\,, \nonumber
		\\
		M_3|_{y\to 0}^{\rm boundary} &= \varepsilon^2 \left[0\right]+\varepsilon^3 \left[\frac{112}{3}\zeta _3+\frac{16}{3}\zeta _2 L_q-\frac{8}{9}L_q^3\right]+\mathcal{O}(\varepsilon^4) \,,
\end{align}
where $L_q=\log q$. It is interesting to note that the two terms in Eq.~\eqref{eq:M3newdef} both contain order $\varepsilon^2$ terms. However, they cancel each other, and $M_3$ starts at order $\varepsilon^3$.

\subsection{Results}
\label{scet:resultb}

\begin{figure}[t!]
  \centering
  \includegraphics[width=7.5cm]{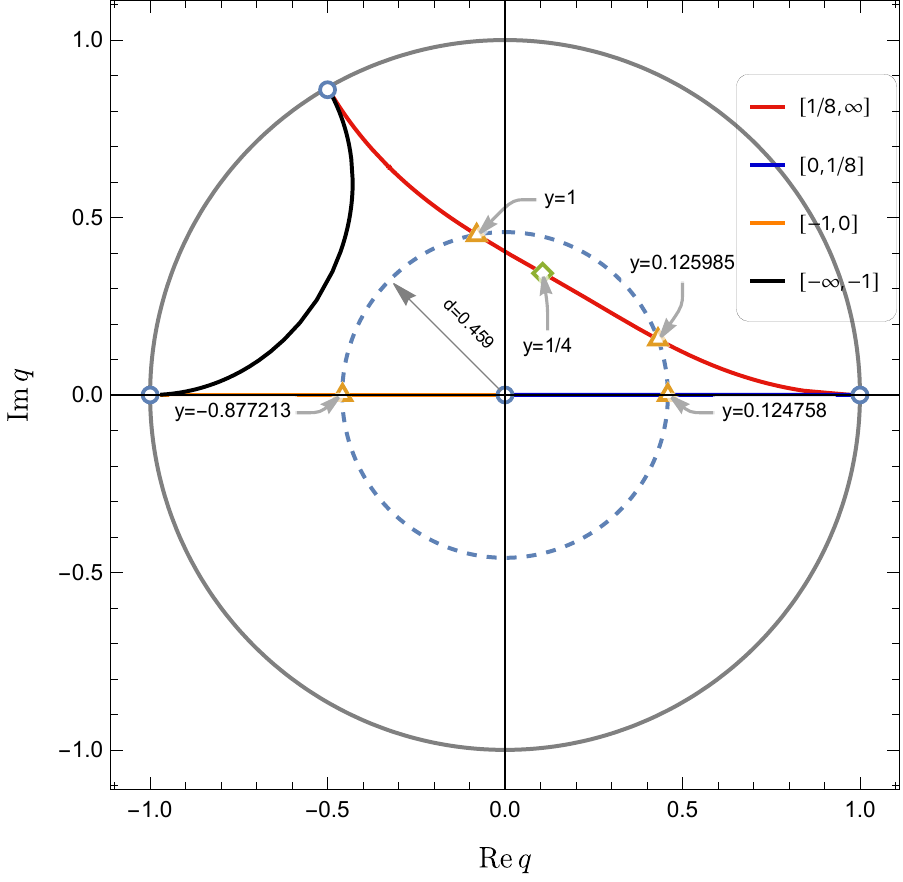}
  \caption{\label{fig:convergentb}Radius of convergence in $q$-space for the series expansion around $q=0$ for family (b), and the corresponding values of $y$.}
\end{figure}

We can express the solutions of differential equations order-by-order in $\varepsilon$ via iterated integrals. It turns out that the leading term in $M_3$ is of order $\varepsilon^3$, i.e., the weight-2 part in $M_3$ originates entirely from sub-sectors and is canceled by the subtraction in Eq.~\eqref{eq:M3newdef}. The leading terms of the three top-sector integrals are given by
\begin{align}
    \label{eq:M1M2M3_family_b}
		M_3 &= \varepsilon^3\bigg[ \frac{112}{3}\zeta_{3} +\zeta_{2}\, I(6\varpi+4\vartheta+3\eta_{3,2};\,i\infty, \tau)-\frac{27}{8}I(\varpi,\eta_{2,2},\eta_{3,2};\, i\infty, \tau) \nonumber
		\\
		&-\frac{9}{8}I(\vartheta,\vartheta,\eta_{2,2};\, i\infty, \tau)-\frac{9}{2} I\left(\eta_{3,2}, \eta_{2,2}, \eta_{2,2}+\frac{9}{16}\eta_{3,2};\,i\infty, \tau\right) \bigg] + \mathcal{O}(\varepsilon^4) \,, \nonumber
		\\
		M_2 &= \varepsilon^3\bigg[ -20\zeta_3 +  \zeta_2\, I(6\eta_{2,3}-\rho;\,i\infty, \tau)  + \frac{9}{32}I\left(\rho, \vartheta, \eta_{2,2};\,i\infty, \tau\right) \nonumber
		\\
		&-9 I\left(\eta_{2,3}, \eta_{2,2}, \eta_{2,2}+\frac{9}{16}\eta_{3,2};\,i\infty, \tau\right) \bigg] + \mathcal{O}(\varepsilon^4) \,, \nonumber
		\\
		M_1 &= \varepsilon^4\bigg[-\frac{31\zeta_4}{2} -20\zeta_3\ln q +  \zeta_2\, I(1, 6\eta_{2,3}-\rho;\,i\infty, \tau)+ \frac{9}{32}I\left(1, \rho, \vartheta, \eta_{2,2};\,i\infty, \tau\right) \nonumber \\
		&-9 I\left(1, \eta_{2,3}, \eta_{2,2}, \eta_{2,2}+\frac{9}{16}\eta_{3,2};\,i\infty, \tau\right) \bigg] + \mathcal{O}(\varepsilon^5) \,.
\end{align}
Like family (a), the combination of $\varepsilon$-factorized differential equations and the UT boundary conditions tells us that the solutions are UT or pure functions in the elliptic sense by the definition of \cite{Frellesvig:2023iwr}.

The convergence of the $q$-expansion, in this case, is again spoiled by the singularities in the sub-sector dependence. For the expansion around $q=0$, the radius of convergence is fixed by the nearest singularity at $y = 1$.~\footnote{The singularity at $y=1/4$ again only affects the sub-sector integral $M_9$.} We depict the convergence region as the disk bounded by the dashed circle in Fig.~\ref{fig:convergentb}. This dashed circle intersects with the $q$-path at four points when varying $y$ from $-\infty+i0$ to $+\infty+i0$. The corresponding values of $y$ at the intersection points can be obtained by combining Eq.~\eqref{eq:newtaub} and the equation of the circle.
In terms of these values, the convergent regions correspond to $y \in (-0.8772,0.1248)\cup (0.1260,1)$, i.e., $x \in (-\infty,-8.013)\cup(-7.937,-1)\cup(1.140,+\infty)$. Note that $y=1/8$ corresponds to a cusp point, and its small neighborhood should be removed, see Eq.~(\ref{eq:cuspsb}) and Fig.~\ref{fig:periodsb}. Note that the convergence is relatively slower when the value of $q$ is close to the boundary circle. Hence, in the following we numerically study the region $x \in (-\infty, -10)\cup (2, \infty)$.
\begin{figure}[!htp]
  \centering
  \includegraphics[width=0.45\textwidth]{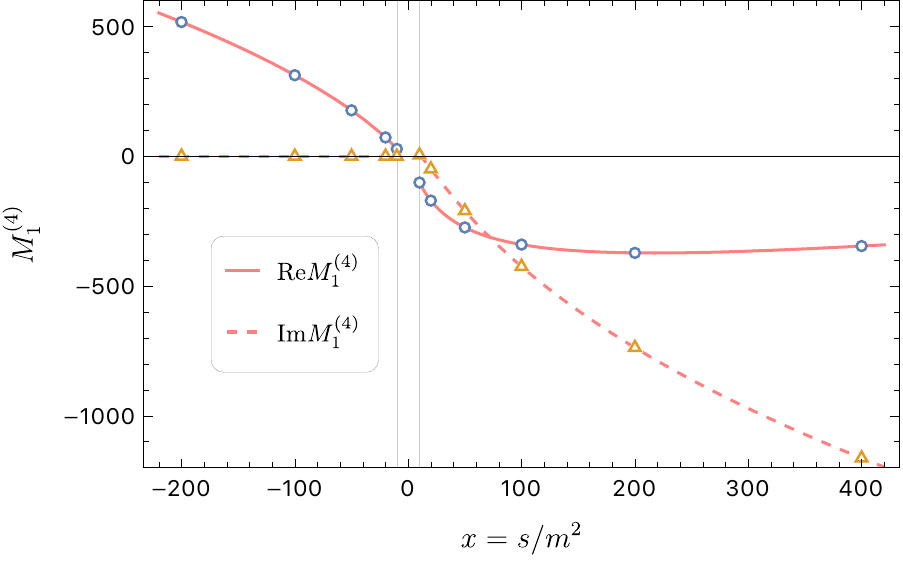}
  \hspace{0.5cm}
  \includegraphics[width=0.45\textwidth]{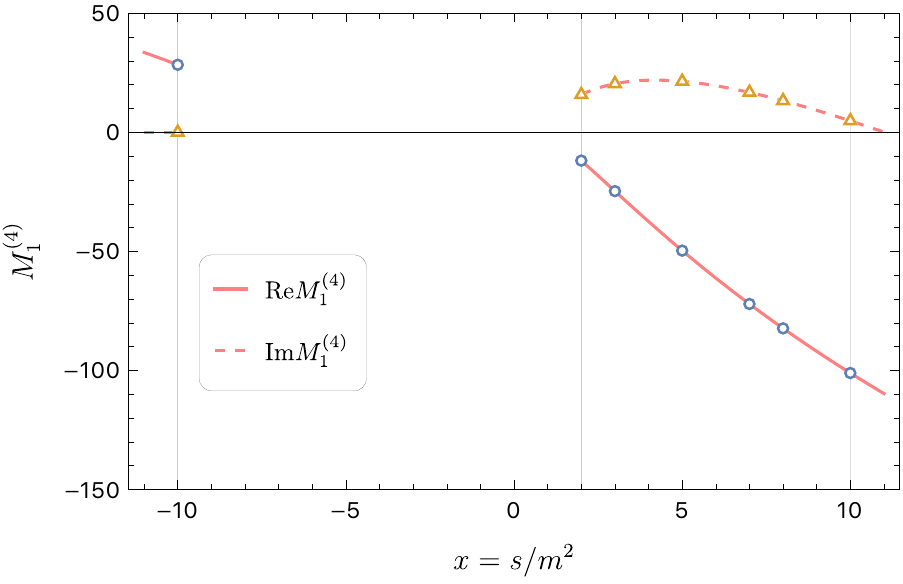}
  \\  
  \includegraphics[width=0.45\textwidth]{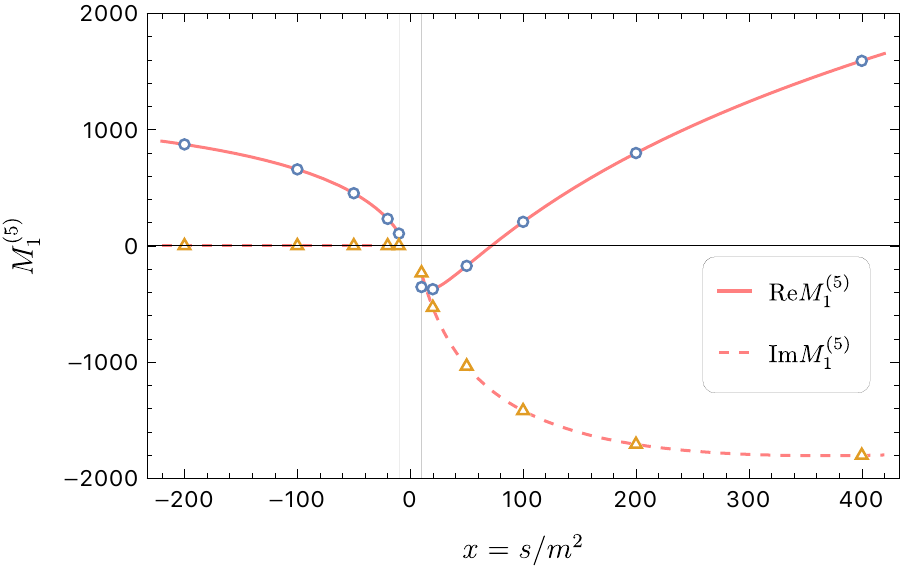}
  \hspace{0.5cm}
  \includegraphics[width=0.45\textwidth]{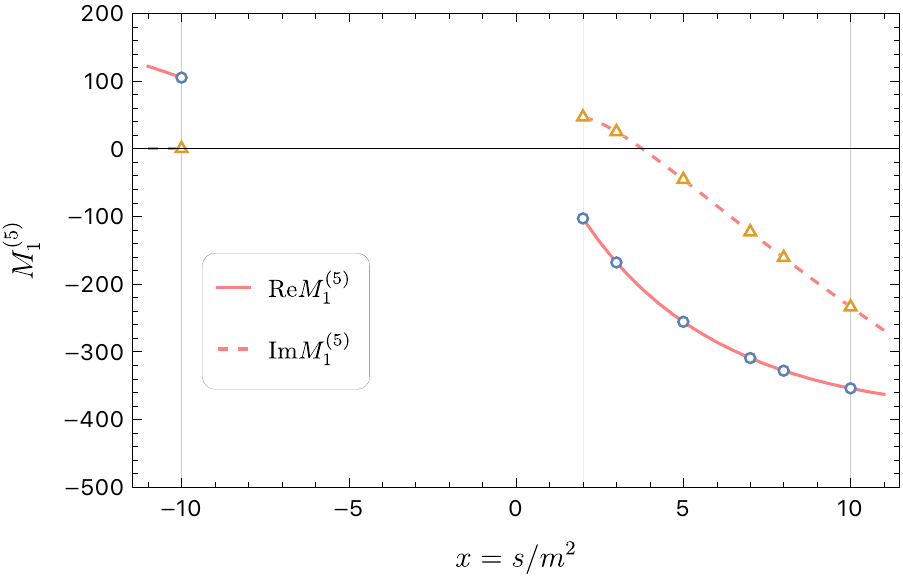}
  \\
  \includegraphics[width=0.45\textwidth]{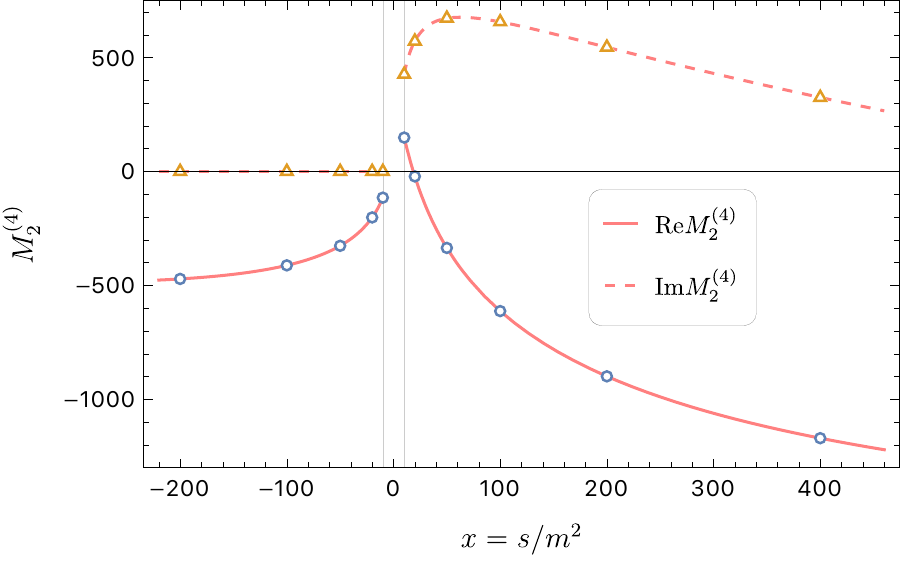}
  \hspace{0.5cm}
  \includegraphics[width=0.45\textwidth]{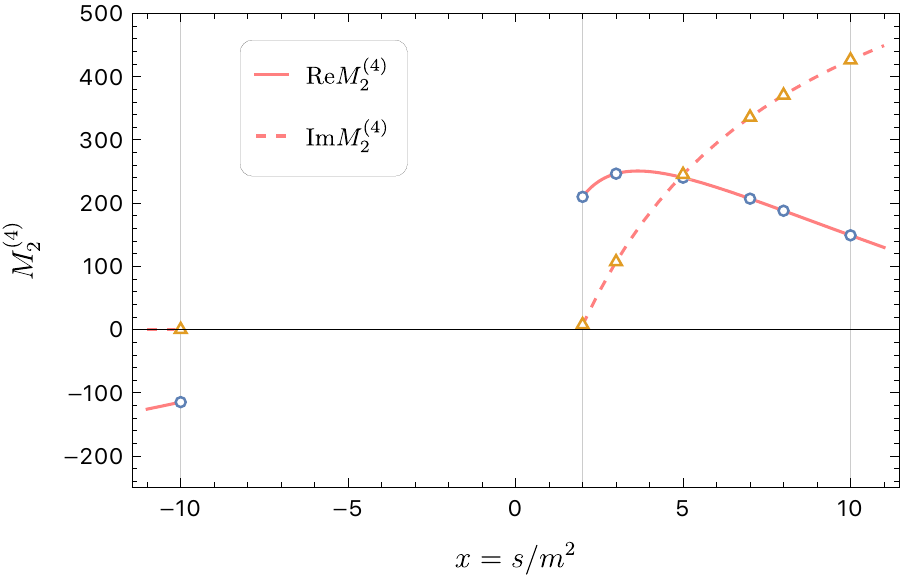}
  \\
  \includegraphics[width=0.45\textwidth]{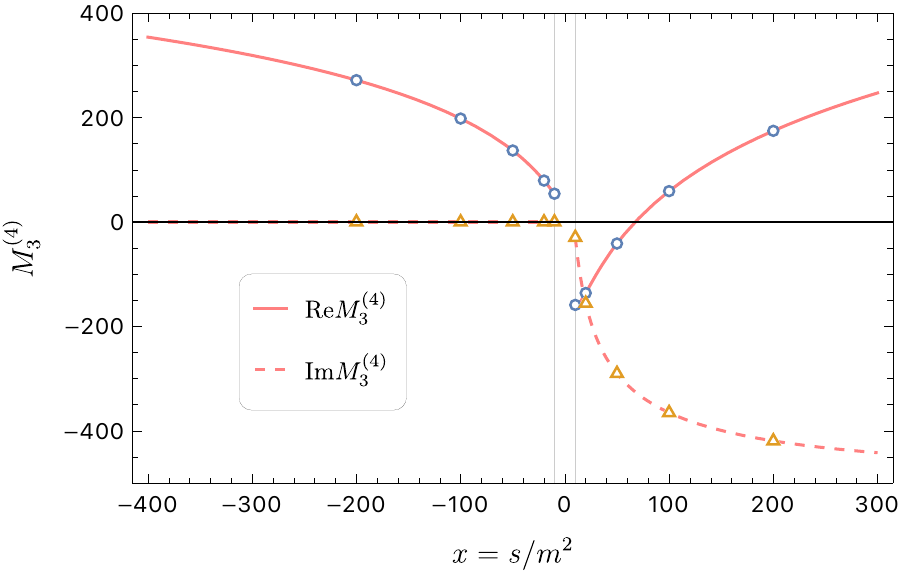}
  \hspace{0.5cm}
  \includegraphics[width=0.45\textwidth]{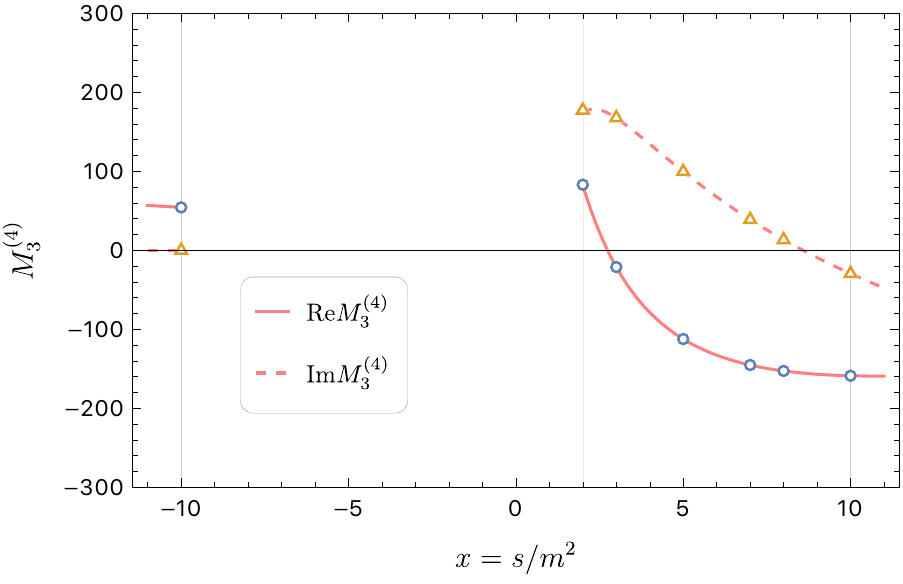}
  \caption{\label{fig:M1w4}Numeric results in family (b) from $q$-expansion (solid and dashed lines) for the weight-4 and weight-5 parts of $M_1$, and the weight-4 parts of $M_2$ and $M_3$. Left: a broad range for $|x| > 10$; right: a closer look at $2 < x < 10$. The results are in good agreement with those from \texttt{AMFlow} (circles and triangles).}
\end{figure}

In Fig.~\ref{fig:M1w4}, we depict the numeric results from $q$-expansion (with the first 8 orders) for the weight-5 part of $M_1$ and the weight-4 parts of $M_{2}$ and $M_{3}$. We compare them with numerical points calculated by \texttt{AMFlow} and find good agreements. For example, there is a relative error at least $\mathcal{O}(10^{-8})$ at $x=-20$ and a relative error at least $\mathcal{O}(10^{-10})$ at $x=20$ for all the functions we plot here. Again, we leave the transformation of the iterated integrals to other kinematic regions for future work.

\section{Conclusions}
\label{sect:conclusions}

In this paper, we have investigated two Feynman integral families involving elliptic curves. Their top sectors are both two-loop non-planar triangle integrals with different configurations of internal masses. They depend on two mass scales. Hence the normalized integrals depend on only one dimensionless variable. We extend the formalism developed for equal-mass Banana integrals to these cases and derive canonical bases for all sectors that satisfy $\varepsilon$-factorized differential equations. Compared with equal-mass Banana integrals, the major new ingredient is the presence of non-trivial sub-sectors. We construct the sub-sector canonical bases using Baikov representations and solve the differential equations in terms of MPLs. The top-sector integrals are represented by iterated integrals with elliptic kernels. We note that the canonical bases have UT boundary conditions. As a result, the top-sector canonical integrals are pure in the elliptic sense.

Our basic tool to achieve $\varepsilon$-factorization is the analysis of the Picard-Fuchs operators corresponding to the elliptic curve underlying the top-sector integrals. The modular variable $\tau$ defined as the ratio of the two periods of the elliptic curve, plays an essential role here. Utilizing this variable, we show that the Picard-Fuchs operator can be factorized under the maximal cut and for $\varepsilon=0$. Family (b) is a more complicated example with three top-sector MIs, but its essence is still an elliptic curve. Its Picard-Fuchs operator is of order three but can be reduced (when $\varepsilon=0$) to the composition of a first-order and a second-order operator. The further factorization of the second-order operator with the modular variable is then derived with the help of a new ingredient, the so-called ``$Y$''-invariant. We emphasize that the Picard-Fuchs method is rather generic and can be applied to geometric objects beyond the elliptic curves, such as those appearing in three- and higher-loop equal-mass Banana integrals~\cite{Pogel:2022yat, Pogel:2022ken, Pogel:2022vat}. Generally speaking, the reason is that fundamental degrees of freedom related to such geometric objects behind Feynman integrals are the periods that are controlled by the Picard-Fuchs operator.

After achieving $\varepsilon$-factorization within the top sectors, we deal with the dependence of the top-sector MIs on the sub-sectors. For family (a) with a simpler sub-sector structure, we find that this dependence is automatically $\varepsilon$-factorized (which is similar to the cases in Banana integrals). This is, however, not true for family (b). We introduce a systematic subtraction scheme to make the sub-sector dependence $\varepsilon$-factorized as well. Family (b) serves as a prototype of a large class of elliptic Feynman integrals, which involve higher-order Picard-Fuchs operators and complicated sub-sectors. We believe that our treatment for family (b) paves the way to deal with those situations.

We investigate in detail the letters (i.e., kernels in the iterated integrals) in the two examples. Since the essential geometric objects are elliptic curves, it is natural to work in the modular space. We find that the letters in the top sector of the family (a) are modular forms of $\Gamma_1(4)$, while those for the family (b) are modular forms of $\Gamma_1(6)$. Letters within sub-sectors themselves are algebraic that can be rationalized with proper variable changes. The complication arises from the nontrivial mixing between top-sector MIs and sub-sector MIs. The letters are combinations of modular forms of $\tau$ and algebraic functions of $y$. The solutions are formally expressed in terms of iterated integrals, which are used to obtain numerical results via $q$-expansion. The evaluation is efficient and agrees perfectly with \texttt{AMFlow}. Due to the singular points in the sub-sector letters, the $q$-expansion around $y=0$ has a finite convergence radius. Extending to other kinematic regions can be achieved by suitable transformations of the iterated integrals. We leave this to future investigations.

\subsection*{Acknowledgements}

We thank Stefan Weinzierl and Sebastian P\"ogel for detailed comments on the manuscript and fruitful discussions, Guo-Xing Wang for discussions about the Mellin-Barnes representations, and Roman Lee for providing \texttt{LiteRed2}. X.W is grateful for the inspiring and fruitful discussion with Christoph Nega and Lorenzo Tancredi during the work.
This work was partly supported by the National Natural Science Foundation of China under Grant No.~11975030 and 12147103, and the Fundamental Research Funds for the Central Universities. X.W was supported by the Excellence Cluster ORIGINS funded by the Deutsche Forschungsgemeinschaft (DFG, German Research Foundation) under Grant No.~EXC - 2094 - 390783311.

\begin{appendix}

\section{Boundary conditions for sub-sector integrals}\label{app:boundary}

This Appendix gives the asymptotic expressions of sub-sector integrals in the limit $y \to 0$ for families (a) and (b).  While the full $\varepsilon$-dependence can be obtained, for this work, it is enough to have the expansion up to weight 5. For family (a), we have
\begin{align}
	M_{3}^{(a)}|^{\rm boundary}_{y\to 0}&=\epsilon^4\left[\frac{\log^4y}{12}-4\zeta_{3}\log y-\frac{\pi^4}{15}\right]+\epsilon^5\left[\frac{\log^5y}{20}-\frac{\pi^2}{18}\log^3y-3\zeta_{3}\log^2y+\frac{2\pi^4}{45}\log y \right.\nonumber \\
	&\left.+\frac{2\pi^2}{3}\zeta_{3}+18\zeta_{5}\right]+\mathcal{O}(\epsilon^6,y) \, , \nonumber \\
	M_{4}^{(a)}|^{\rm boundary}_{y\to 0}&=\epsilon^4\left[\frac{\pi^2}{12}\log^2y+\frac{7\pi^4}{360}\right]+\epsilon^5\left[\frac{\pi^2}{12}\log^3y+\frac{\zeta_{3}}{2}\log^2y+\frac{\pi^4}{90}\log y -\frac{\pi^2}{3}\zeta_{3}+\zeta_{5}\right] \nonumber \\
	&+\mathcal{O}(\epsilon^6,y) \, , \nonumber \\
	M_{5}^{(a)}|^{\rm boundary}_{y\to 0}&=\epsilon^2\left[\frac{\log^2y}{2}+\frac{\pi^2}{6}\right]+\epsilon^3\left[\frac{2\log^3y}{3}+\frac{\pi^2}{3}\log y-4\zeta_{3}\right] \nonumber \\
	&+\epsilon^4\left[\frac{5\log^4y}{12}+\frac{\pi^2}{4}\log^2y-8\zeta_{3}\log y\right]+\epsilon^5\left[\frac{11\log^5y}{60}+\frac{\pi^2}{6}\log^3y-\frac{31\zeta_{3}}{3}\log^2y \right. \nonumber \\
	&\left.-\frac{17\pi^4}{180}\log y-\frac{13}{9}\pi^2\zeta_{3}-32\zeta_{5}\right]+\mathcal{O}(\epsilon^6,y) \, , \nonumber \\
	M_{6}^{(a)}|^{\rm boundary}_{y\to 0}&=\epsilon^3\left[-\frac{\log^3y}{6}-\frac{\pi^2}{6}\log y+2\zeta_{3}\right]+\epsilon^4\left[-\frac{\log^4y}{8}-\frac{\pi^2}{12}\log^2y+2\zeta_{3}\log y-\frac{\pi^4}{72}\right] \nonumber \\
	&+\epsilon^5\left[-\frac{7}{120}\log^5y-\frac{\pi^2}{18}\log^3y+3\zeta_{3}\log^2y+\frac{\pi^4}{72}\log y+\frac{2\pi^2}{3}\zeta_{3}+14\zeta_{5}\right]+\mathcal{O}(\epsilon^6,y) \, , \nonumber \\
	M_{7}^{(a)}|^{\rm boundary}_{y\to 0}&=\epsilon^3\left[-\frac{\log^3y}{3}+4\zeta_{3}\right]+\epsilon^4\left[-\frac{\log^4y}{4}+\frac{\pi^2}{6}\log^2y+6\zeta_{3}\log y-\frac{2\pi^4}{45}\right] \nonumber \\
	&+\epsilon^5\left[-\frac{7}{60}\log^5y+\frac{\pi^2}{9}\log^3y+8\zeta_{3}\log^2y+\frac{13\pi^4}{180}\log y-\frac{\pi^2}{3}\zeta_{3}+30\zeta_{5}\right]+\mathcal{O}(\epsilon^6,y) \, , \nonumber \\
	M_{8}^{(a)}|^{\rm boundary}_{y\to 0}&=\epsilon^2\left[\log^2y\right]+\epsilon^3\left[\log^3y-\frac{\pi^2}{3}\log y-6\zeta_{3}\right]+\epsilon^4\left[\frac{7}{12}\log^4y-\frac{\pi^2}{3}\log^2y-16\zeta_{3}\log y\right. \nonumber \\
	&\left.-\frac{13\pi^4}{180} \right]+\epsilon^5\left[\frac{\log^5y}{4}-\frac{2\pi^2}{9}\log^3y-\frac{56}{3}\zeta_{3}\log^2y-\frac{\pi^4}{4}\log y+\frac{5\pi^2}{3}\zeta_{3}-42\zeta_{5} \right] \nonumber \\
	&+\mathcal{O}(\epsilon^6,y) \, , \nonumber \\
	M_{9}^{(a)}|^{\rm boundary}_{y\to 0}&=\epsilon\left[-\log y\right]+\epsilon^2\left[-2\log^2y+\frac{\pi^2}{6}\right]+\epsilon^3\left[-\frac{5}{3}\log^3y+\frac{\pi^2}{2}\log y+11\zeta_{3}\right]\nonumber \\
	&+\epsilon^4\left[-\frac{11}{12}\log^4y+\frac{\pi^2}{2}\log^2y+\frac{80}{3}\zeta_{3}\log y +\frac{29\pi^4}{180}\right]+\epsilon^5\left[-\frac{23}{60}\log^5y+\frac{\pi^2}{3}\log^3y\right. \nonumber \\
	&\left.+\frac{88}{3}\zeta_{3}\log^2y+\frac{49\pi^4}{120}\log y-\frac{47}{18}\pi^2\zeta_{3}+69\zeta_{5} \right]+\mathcal{O}(\epsilon^6,y) \, , \nonumber \\
	M_{10}^{(a)}|^{\rm boundary}_{y\to 0}&=-1+\epsilon\left[-\log y\right]+\epsilon^2\left[-\frac{1}{2}\log^2y\right]+\epsilon^3\left[-\frac{1}{6}\log^3y+\frac{8}{3}\zeta_3\right]+\epsilon^4\left[-\frac{1}{24}\log^4y \right. \nonumber \\
	&\left.+\frac{8}{3}\zeta_3\log y+\frac{\pi^4}{30}\right]+\epsilon^5\left[-\frac{1}{120}\log^5y+\frac{4}{3}\zeta_3\log^2y+\frac{\pi^4}{30}\log y+\frac{32}{5}\zeta_5\right]+\mathcal{O}(\epsilon^6,y) \, , \nonumber \\
	M_{11}^{(a)}|^{\rm boundary}_{y\to 0}&=1+\epsilon^2\left[\frac{\pi^2}{6}\right]+\epsilon^3\left[-\frac{2}{3}\zeta_3\right]+\epsilon^4\left[\frac{7\pi^4}{360}\right]+\epsilon^5\left[-\frac{\pi^2}{9}\zeta_3-\frac{2}{5}\zeta_5\right]+\mathcal{O}(\epsilon^6,y) \, .
	\label{eq:family_a_bd_sub}
\end{align}
and for family (b) we have
\begin{align}
		M_{4}^{(b)}|^{\rm boundary}_{y\to 0}&=\epsilon^4\left[\frac{\pi^2}{12}\log^2y-\zeta_{3}\log y +\frac{7\pi^4}{360}\right]+\epsilon^5\left[\frac{\pi^2}{12}\log^3y-\frac{\pi^4}{360}\log y-\frac{2}{3}\pi^2\zeta_3+9\zeta_5\right] \nonumber \\
		&+\mathcal{O}(\epsilon^6,y) \, , \nonumber \\
		M_{5}^{(b)}|^{\rm boundary}_{y\to 0}&=\epsilon\left[-\frac{1}{2}\log y\right]+\epsilon^2\left[-\log^2y\right]+\epsilon^3\left[-\frac{5}{6}\log^3y+\frac{5\pi^2}{12}\log y+4\zeta_3\right] \nonumber \\
		&+\epsilon^4\left[-\frac{11}{24}\log^4y+\frac{7\pi^2}{12}\log^2y +\frac{31}{3}\zeta_3\log y+\frac{17\pi^4}{360}\right]+\epsilon^5\left[-\frac{23}{120}\log^5y \right.\nonumber \\
		&\left.+\frac{4\pi^2}{9}\log^3y+\frac{44}{3}\zeta_3\log^2y+\frac{83\pi^4}{720}\log y -\frac{4\pi^2}{3}\zeta_3+29\zeta_5\right]+\mathcal{O}(\epsilon^6,y) \, , \nonumber \\
		M_{6}^{(b)}|^{\rm boundary}_{y\to 0}&=\epsilon^4\left[\frac{\pi^2}{12}\log^2y\!-\!2\zeta_3\log y+\frac{11\pi^4}{180}\right]\!+\!\epsilon^5\left[\frac{\pi^2}{12}\log^3y\!-\!\frac{\zeta_3}{2}\log^2y\!-\!\frac{\pi^4}{18}\log y \right. \nonumber \\
		&\left.+\frac{5\pi^2}{6}\zeta_3+5\zeta_5 \right]\!+\!\mathcal{O}(\epsilon^6,y) , \nonumber \\
		M_{7}^{(b)}|^{\rm boundary}_{y\to 0}&=\epsilon^3\left[\frac{1}{6}\log^3y+\frac{\pi^2}{3}\log y+2\zeta_3\right]+\epsilon^4\left[-\frac{\pi^2}{4}\log^2y-\zeta_3\log y-\frac{41\pi^4}{180}\right] \nonumber \\
		&+\epsilon^5\left[\frac{1}{120}\log^5y+\frac{\pi^2}{12}\log^3y-\frac{\zeta_3}{2}\log^2y+\frac{19\pi^4}{72}\log y+\frac{\pi^2}{6}\zeta_3+10\zeta_5\right] +\mathcal{O}(\epsilon^6,y) \, , \nonumber \\
		M_{8}^{(b)}|^{\rm boundary}_{y\to 0}&=\epsilon^4\left[\frac{1}{8}\log^4y+\frac{\pi^2}{4}\log^2y-\zeta_3\log y+\frac{\pi^4}{30}\right]+\epsilon^5\left[\frac{1}{20}\log^5y-\frac{\pi^2}{9}\log^3y-\frac{5\zeta_3}{2}\log^2y \right. \nonumber \\
		&\left.-\frac{71\pi^4}{360}\log y-\frac{\pi^2}{2}\zeta_3-\zeta_5\right]+\mathcal{O}(\epsilon^6,y) \, , \nonumber \\
		M_{9}^{(b)}|^{\rm boundary}_{y\to 0}&=\epsilon^2\left[-\frac{1}{2}\log^2y-\frac{\pi^2}{6}\right]+\epsilon^3\left[-\frac{2}{3}\log^3y-\frac{\pi^2}{3}\log y+4\zeta_3\right]\!+\!\epsilon^4\!\left[-\frac{5}{12}\log^4y\right. \nonumber \\
		&\left.-\frac{\pi^4}{4}\log^2y+8\zeta_3\log y\right] +\epsilon^5\left[-\frac{11}{60}\log^5y-\frac{\pi^2}{6}\log^3y+\frac{31}{3}\zeta_3\log^2y+\frac{17\pi^4}{180}\log y \right. \nonumber \\
		&\left.+\frac{13\pi^2}{9}\zeta_3+32\zeta_5\right]+\mathcal{O}(\epsilon^6,y) \, .
		\nonumber \\
		M_{10}^{(b)}|^{\rm boundary}_{y\to 0}&=\epsilon^3\left[-\frac{1}{6}\log^3y-\frac{\pi^2}{6}\log y+2\zeta_3\right]+\epsilon^4\left[-\frac{1}{8}\log^4y-\frac{\pi^2}{12}\log^2y+2\zeta_3\log y-\frac{\pi^4}{72}\right] \nonumber \\
		&+\epsilon^5\left[-\frac{7}{120}\log^5y-\frac{\pi^2}{18}\log^3y+3\zeta_3\log^2y+\frac{\pi^4}{72}\log y+\frac{2\pi^2}{3}\zeta_3+14\zeta_5\right]+\mathcal{O}(\epsilon^6,y) \, , \nonumber \\
		M_{11}^{(b)}|^{\rm boundary}_{y\to 0}&=1\!+\epsilon^2\!\left[-\frac{1}{2}\log^2y\right]\!+\epsilon^3\!\left[-\frac{5}{6}\log^3y-\frac{\pi^2}{6}\log y+\frac{10}{3}\zeta_3\right]\!+\!\epsilon^4\!\left[-\frac{13}{24}\log^4y+\frac{\pi^2}{6}\log^2y\right. \nonumber \\
		&\biggl.+10\zeta_3\log y \biggr] +\epsilon^5\left[-\frac{29}{120}\log^5y+\frac{\pi^2}{6}\log^3y+\frac{46}{3}\zeta_3\log^2y+\frac{53\pi^4}{360}\log y \right. \nonumber \\
		&\left.+\frac{\pi^2}{3}\zeta_3+\frac{198}{5}\zeta_5\right]+\mathcal{O}(\epsilon^6,y) \, , \nonumber \\
		M_{12}^{(b)}|^{\rm boundary}_{y\to 0}&=\epsilon\left[\frac{1}{2}\log y\right]+\epsilon^2\left[\frac{3}{2}\log^2y+\frac{\pi^2}{6}\right]+\epsilon^3\left[\frac{3}{2}\log^3y-\frac{\pi^2}{12}\log y-7\zeta_3\right]+\epsilon^4\left[\frac{7}{8}\log^4y \right. \nonumber \\
		&\left.-\frac{\pi^2}{3}\log^2y -\frac{61}{3}\zeta_3\log y-\frac{23\pi^4}{360}\right]\!+\epsilon^5\!\left[\frac{3}{8}\log^5y-\frac{5\pi^2}{18}\log^3y \right. \nonumber \\
		&\left.-26\zeta_3\log^2y-\frac{227\pi^4}{720}\log y +\frac{13\pi^2}{18}\zeta_3\!-\!59\zeta_5\right]+\mathcal{O}(\epsilon^6,y) \, , \nonumber \\
		M_{13}^{(b)}|^{\rm boundary}_{y\to 0}&=-\frac{1}{4}+\epsilon\left[-\frac{1}{2}\log y\right]+\epsilon^2\left[-\frac{1}{2}\log^2y-\frac{\pi^2}{24}\right]+\epsilon^3\left[-\frac{1}{3}\log^3y-\frac{\pi^2}{12}\log y+\frac{13}{6}\zeta_3\right] \nonumber \nonumber \\
		&+\epsilon^4\left[-\frac{1}{6}\log^4y -\frac{\pi^2}{12}\log^2y+\frac{13}{3}\zeta_3\log y+\frac{41\pi^4}{1440}\right]+\epsilon^5\left[-\frac{1}{15}\log^5y-\frac{\pi^2}{18}\log^3y\right. \nonumber \\
		&\left.+\frac{13}{3}\zeta_3\log^2y+\frac{41\pi^4}{720}\log y +\frac{13\pi^2}{36}\zeta_3+\frac{121}{10}\zeta_5\right]+\mathcal{O}(\epsilon^6,y) \, , \nonumber \\
		M_{14}^{(b)}|^{\rm boundary}_{y\to 0}&=M_{10}^{(a)}|^{\rm boundary}_{y\to 0} \, , \nonumber \\
		M_{15}^{(b)}|^{\rm boundary}_{y\to 0}&=-\frac{1}{2}+\epsilon^2\left[-\frac{\pi^2}{4}\right]+\epsilon^3\left[\frac{4}{3}\zeta_3\right]+\epsilon^4\left[-\frac{7\pi^4}{80}\right]+\epsilon^5\left[\frac{2\pi^2}{3}\zeta_3+\frac{16}{5}\zeta_5\right]++\mathcal{O}(\epsilon^6,y) \, , \nonumber \\
		M_{16}^{(b)}|^{\rm boundary}_{y\to 0}&=\epsilon\left[-\frac{1}{2}\log y\right]+\epsilon^2\left[-\frac{3}{4}\log^2t+\frac{\pi^2}{12}\right]+\epsilon^3\left[-\frac{7}{12}\log^3y+\frac{\pi^2}{6}\log y+4\zeta_3\right] \nonumber \\
		&+\epsilon^4\left[-\frac{5}{16}\log^4y +\frac{\pi^2}{6}\log^2y+\frac{28}{3}\zeta_3\log y+\frac{\pi^4}{16}\right]+\epsilon^5\left[-\frac{31}{240}\log^5y+\frac{\pi^2}{9}\log^3y \right. \nonumber \\
		&\left. +10\zeta_3\log^2y+\frac{17\pi^4}{120}\log y -\frac{8\pi^2}{9}\zeta_3+24\zeta_5\right]+\mathcal{O}(\epsilon^6,y) \, , \nonumber \\
		M_{17}^{(b)}|^{\rm boundary}_{y\to 0}&=\frac{1}{4}+\epsilon\left[\log y\right]+\epsilon^2\left[\frac{5}{4}\log^2y-\frac{\pi^2}{8}\right]+\epsilon^3\left[\frac{11}{12}\log^3y-\frac{\pi^2}{4}\log y-\frac{20\zeta_3}{3}\right] \nonumber \\
		&+\epsilon^4\left[\frac{23}{48}\log^4y -\frac{\pi^2}{4}\log^2y-\frac{44}{3}\zeta_3\log y-\frac{49\pi^4}{480}\right]+\epsilon^5\left[\frac{47}{240}\log^5y-\frac{\pi^2}{6}\log^3y \right. \nonumber \\
		&\left.-\frac{46}{3}\zeta_3\log^2y-\frac{53\pi^4}{240}\log y +\frac{4\pi^2}{3}\zeta_3-\frac{188}{5}\zeta_5\right]+\mathcal{O}(\epsilon^6,y) \, , \nonumber \\
		M_{18}^{(b)}|^{\rm boundary}_{y\to 0}&=M_{11}^{(a)}|^{\rm boundary}_{y\to 0} \, .
\end{align}

\end{appendix}
\bibliographystyle{JHEP}
\bibliography{master.bib}

\end{document}